\documentclass[twocolumn,preprintnumbers,amsmath,amssymb,nobibnotes,nofootinbib,superscriptaddress]{revtex4}
\usepackage{graphicx}
\usepackage{dcolumn}
\usepackage[usenames,dvipsnames]{xcolor}
\RequirePackage[colorlinks=true
,urlcolor=blue
,anchorcolor=blue
,citecolor=blue
,filecolor=blue
,linkcolor=blue
,menucolor=blue
,pagecolor=blue
,linktocpage=true
,pdfproducer=medialab
]{hyperref}
\usepackage{amsmath}
\usepackage{mathrsfs}
\usepackage{slashed}
\usepackage{cancel}

\newcommand{\LL}{\mathscr{L}}
\newcommand{\cG}{\mathscr{G}}

\def\cD{{\cal D}}
\def\cF{{\cal F}}

\def\cG{{\cal G}}
\def\cGSM{{\cal G}_{SM}}

\def\cO{{\cal O}}
\def\cOL{{\cal O}_L}
\def\cOR{{\cal O}_R}
\def\cOLR{{\cal O}_{L(R)}}
\def\cOLt{\widetilde{{\cal O}}_L}
\def\cORt{\widetilde{{\cal O}}_R}
\def\cOLRt{\widetilde{{\cal O}}_{L(R)}}
\def\cP{{\cal P}}

\def\Tr{{\rm Tr}}

\def\be{\begin{equation}}
\def\ee{\end{equation}}
\def\beq{\begin{equation}}
\def\eeq{\end{equation}}
\def\bc{\begin{center}}
\def\ec{\end{center}}
\def\bea{\begin{eqnarray}}
\def\eea{\end{eqnarray}}

\def\nt{\noindent}
\newcommand{\mean}[1]{\langle#1\rangle}
\newcommand{\derp}{\partial}
\newcommand{\LT}{\mathbf{L}}
\newcommand{\RT}{\mathbf{R}}
\newcommand{\UY}{\mathbf{U_Y}}
\newcommand{\UH}{\mathbf{U}}
\newcommand{\UHL}{\mathbf{U}_L}
\newcommand{\UHR}{\mathbf{U}_R}
\newcommand{\UHLR}{\mathbf{U}_{L(R)}}

\newcommand{\TL}{\mathbf{T}}
\newcommand{\VL}{\mathbf{V}}
\newcommand{\DL}{D}

\newcommand{\DLL}{\mathcal{D}}

\newcommand{\TLL}{\mathbf{T}_L}
\newcommand{\TLR}{\mathbf{T}_R}
\newcommand{\VLLmuu}{\mathbf{V}^\mu_L}
\newcommand{\VLLmud}{\mathbf{V}_{\mu,\,L}}
\newcommand{\VLLnuu}{\mathbf{V}^\nu_L}

\newcommand{\VLRmuu}{\mathbf{V}^\mu_R}
\newcommand{\VLRmud}{\mathbf{V}_{\mu,\,R}}
\newcommand{\VLRnuu}{\mathbf{V}^\nu_R}

\newcommand{\VLLmuut}{\widetilde{\mathbf{V}}^\mu_L}
\newcommand{\VLLmudt}{\widetilde{\mathbf{V}}_{\mu,\,L}}
\newcommand{\VLLnuut}{\widetilde{\mathbf{V}}^\nu_L}
\newcommand{\VLLnudt}{\widetilde{\mathbf{V}}_{\nu,\,L}}
\newcommand{\VLRmuut}{\widetilde{\mathbf{V}}^\mu_R}
\newcommand{\VLRmudt}{\widetilde{\mathbf{V}}_{\mu,\,R}}
\newcommand{\VLRnuut}{\widetilde{\mathbf{V}}^\nu_R}
\newcommand{\VLRnudt}{\widetilde{\mathbf{V}}_{\nu,\,R}}

\newcommand{\BBu}{B^{\mu\nu}}
\newcommand{\BBd}{B_{\mu\nu}}

\newcommand{\WWLu}{W^{\mu\nu}_L}
\newcommand{\WWRu}{W^{\mu\nu}_R}

\newcommand{\WWLut}{\widetilde{W}^{\mu\nu}_L}
\newcommand{\WWRut}{\widetilde{W}^{\mu\nu}_R}

\newcommand{\WWRdt}{\widetilde{W}_{\mu\nu,\,R}}

\newcommand{\WWWLut}{\widetilde{W}^{\rho\sigma}_L}
\newcommand{\WWWRut}{\widetilde{W}^{\rho\sigma}_R}

\newcommand{\TLchi}{\mathbf{T}_\chi}
\newcommand{\VLchimuu}{\mathbf{V}^\mu_\chi}
\newcommand{\VLchimud}{\mathbf{V}_{\mu,\,\chi}}
\newcommand{\VLchinuu}{\mathbf{V}^\nu_\chi}
\newcommand{\VLchinud}{\mathbf{V}_{\nu,\,\chi}}

\newcommand{\WWchiu}{W^{\mu\nu}_\chi}

\newcommand{\WWchiut}{\widetilde{W}^{\mu\nu}_\chi}
\newcommand{\WWWchiu}{W^{\rho\sigma}_\chi}

\newcommand{\TLLR}{\mathbf{T}_{L(R)}}
\newcommand{\VLLRmuu}{\mathbf{V}^\mu_{L(R)}}
\newcommand{\VLLRnuu}{\mathbf{V}^\nu_{L(R)}}
\newcommand{\TLLt}{\widetilde{\mathbf{T}}_{L}}
\newcommand{\TLRt}{\widetilde{\mathbf{T}}_{R}}
\newcommand{\TLLRt}{\widetilde{\mathbf{T}}_{L(R)}}
\newcommand{\VLLRmuut}{\widetilde{\mathbf{V}}^\mu_{L(R)}}
\newcommand{\VLRLmuut}{\widetilde{\mathbf{V}}^\mu_{R(L)}}
\newcommand{\VLRLmudt}{\widetilde{\mathbf{V}}_{\mu,R(L)}}
\newcommand{\VLRLnuut}{\widetilde{\mathbf{V}}^\nu_{R(L)}}

\newcommand{\WWLRu}{W^{\mu\nu}_{L(R)}}
\newcommand{\WWLRut}{\widetilde{W}^{\mu\nu}_{L(R)}}
\newcommand{\WWRLut}{\widetilde{W}^{\mu\nu}_{R(L)}}

\newcommand{\fL}{f_L}
\newcommand{\fR}{f_R}
\newcommand{\fLR}{f_{L(R)}}
\newcommand{\fRL}{f_{R(L)}}
\newcommand{\fchi}{f_\chi}
\newcommand{\gL}{g_L}
\newcommand{\gR}{g_R}
\newcommand{\gLR}{g_{L(R)}}

\newcommand{\gchi}{g_\chi}

\newcommand{\tr}{\Tr}

\newcommand{\brown}[1]{\color[rgb]{0.8,0.1,0.} #1 \color{black}}

\begin{document}
%
%

\title{Spin--1 resonances in a non-linear left-right dynamical Higgs context}

\author{Juan Yepes}
\email{juyepes@itp.ac.cn}
\affiliation{State Key Laboratory of Theoretical Physics and Kavli Institute for Theoretical Physics China (KITPC)\\
Institute of Theoretical Physics, Chinese Academy of Sciences, Beijing 100190, P. R. China}

\begin{abstract}

All the possible CP--conserving non--linear operators up to the $p^4$--order in the Lagrangian expansion are clearly and completely listed for the case of an emerging new physics field content in the nature, more specifically, from spin--1 resonances sourced by the straightforward extension of the SM local gauge symmetry to the larger local group $SU(2)_L\otimes SU(2)_R\otimes U(1)_{B-L}$, within a non--linear electroweak chiral context coupled to a light dynamical Higgs. The physical effects induced by integrating out the right handed fields from the physical spectrum are briefly analysed. The relevant set of effective operators have been identified at low energies.

\end{abstract}
\maketitle


\section{Introduction}

\nt The LHC discovery  of a new scalar resonance~\cite{:2012gk,:2012gu} and its experimental confirmation as a particle resembling the Higgs boson~\cite{Englert:1964et,
Higgs:1964ia,Higgs:1964pj} have finally established the Standard Model (SM) as a successful and consistent framework of electroweak symmetry breaking (EWSB). The role of the Higgs particle in the EWSB mechanism signals different BSM scenarios. In one class of models, the Higgs is introduced  as an elementary scalar doublet transforming linearly under the SM gauge group $SU(2)_L \otimes U(1)_Y$. An alternative is  to postulate its nature as  emerging from a given strong dynamics sector at the TeV or slightly higher scale, in which the Higgs enters either as an EW doublet or as a member of other representations: a singlet in all generality.  Both cases call for new physics (NP) around the TeV scale and tend to propose the existence of lighter exotic resonances which have failed to show up in data so far.

The alternative case assumes a non-perturbative Higgs dynamics associated to a strong interacting sector at $\Lambda_s$-scale, with a explicitly non--linear implementation of   the symmetry in the scalar sector. 
 These strong dynamics frameworks all share a reminiscence of the long ago proposed ``Technicolor" formalism \cite{Susskind:1978ms,Dimopoulos:1979es,Dimopoulos:1981xc}, 
  in which no 
Higgs particle was proposed in the low-energy physical spectrum and only three would-be-Goldstone bosons (GB) were present with an associated scale $f$ identified with the electroweak scale $f=v\equiv246$ GeV (respecting $f \ge \Lambda_s/4 \pi$~\cite{Manohar:1983md}), and responsible a posteriori for the weak gauge boson masses. The experimental discovery of a light Higgs boson, not accompanied of extra resonances, has led to a revival in this direction relying on the fact that the Higgs particle $h$ may be light as being itself a GB	 resulting from the spontaneous breaking of a strong dynamics with symmetry group  $G$ at the scale $\Lambda_s$~\cite{Kaplan:1983fs,Kaplan:1983sm,Banks:1984gj,
Georgi:1984ef,Georgi:1984af,Dugan:1984hq}.  A subsequent source of explicit breaking of $G$ would allow the Higgs boson to pick a small mass, much as the pion gets a mass in QCD, and develops a potential with a non-trivial minimum $\langle h\rangle$. Only via this explicit breaking would the EW gauge symmetry be broken and the electroweak scale $v$ be generated, distinct from $f$.   Three scales enter thus into the scenario now: $f$, $v$ and $\langle h\rangle$, although a model-dependent constraint will link them. 
The strength of non--linearity is quantified by a new parameter
\beq
\xi\equiv \frac{v^2}{f^2}\,,
\label{xi}
\eeq

\nt such that, $f\sim v$ ($\xi \sim 1$) characterizes non--linear constructions, whilst $f\gg v$ ($\xi \ll 1$) labels  regimes approaching the linear one. As a result, for non-negligible $\xi$ there may be corrections 
to the size of the SM couplings observable at low energies due to the NP contributions. 

In this work an EW strongly interacting sector coupled to the light Higgs particle will be assumed. Furthermore, motivated by the exciting high energy regimes
reachable at the LHC and future colliders, this work faces the possibility for detecting non--zero signals arising out from some emerging new physics field content in the nature, more specifically, from spin--1 resonances brought into the game via the extension of the SM local gauge symmetry $\cG_{SM}=SU(2)_L\otimes U(1)_Y$ to the larger local group $\cG=SU(2)_L\otimes SU(2)_R\otimes U(1)_{B-L}$ (see~\cite{LRSM1,LRSM2} for left-right symmetric models literature). Low energy effects from such extended gauge field sector can be tackled through a systematic model-independent EFT approach. The idea is to employ a non--linear $\sigma$--model to account for the strong dynamics giving rise to the GB, that is the $W^\pm_L$ and $Z_L$ longitudinal components that leads to introduce the Goldstone scale $\fL$ (basically the scale $f$ in Eq.~\eqref{xi}), together with the corresponding GB from the extended local group, i.e. the additional $W^\pm_R$ and $Z_R$ longitudinal degrees of freedom and the associated Goldstone scale $\fR$. Finally, this non--linear $\sigma$--model effective Lagrangian is coupled a posteriori to a scalar singlet $h$ in a general way through powers of $h/\fL$~\cite{Georgi:1984af}, being the scale suppression dictated by $\fL$, as it is the scale where $h$ is generated as a GB prior to the extension of the SM local group $\cG_{SM}$ to the larger one $\cG$. 

In this work it is analysed the physical picture of spin--1 resonances dictated by the larger local gauge group $\cG$, with an underlying strong interacting scenario coupled to a light Higgs particle. The non--linear EFT construction assumed here provides the complete tower of pure gauge and gauge-$h$ operators up to the $p^4$--order in the Lagrangian expansion, and restricted only to the CP-conserving bosonic sector. This analysis enlarges and completes the operator basis previously considered in Refs.~\cite{Zhang:2007xy,Wang:2008nk} in the context of left--right symmetric EW chiral models, it generalizes also the work done in Refs.~\cite{Appelquist:1980vg,Longhitano:1980iz,Longhitano:1980tm,
Feruglio:1992wf,Appelquist:1993ka}, and it extends as well Refs.~\cite{Alonso:2012px,Brivio:2013pma} to the case of a larger local gauge symmetry $\cG$ in the context of non--linear EW interactions coupled to a light Higgs particle. It may also be considered as a generic UV completion of the low energy non--linear treatment of Refs.~\cite{Appelquist:1980vg,Longhitano:1980iz,Longhitano:1980tm,
Feruglio:1992wf,Appelquist:1993ka}  and~\cite{Alonso:2012px,Brivio:2013pma}.


\section{Theoretical set--up}
\label{Theoretical-set-up}

\nt The original SM local symmetry $SU(2)_L\otimes U(1)_Y$ is enlarged now to the larger local group $SU(2)_L\otimes SU(2)_R\otimes U(1)_{B-L}$. As soon as the original SM group symmetry $\cG_{SM}$ is extended by calling for the inclusion of the additional local symmetry $SU(2)_R$, the corresponding physical spectrum has to be enlarged as well. In this work, the previous $\cG_{SM}$--gauge field content\footnote{Henceforth the $SU(2)_L$ $W$--field will be labelled with the corresponding suffix $L$.} $B^{\mu}$, $W^{a\mu}_L$ is augmented by the additional $SU(2)_R$ gauge field $W^{a\mu}_R$ leading to have a posteriori the right handed fields $W^{\pm}_R$, $Z_R$ (here labelled with a suffix $R$) and their longitudinal 
degrees of freedom encoded via their corresponding Goldstone bosons fields. The transformation properties of the longitudinal degrees of freedom of the electroweak gauge bosons will be parametrized\footnote{Usually  for the case of a purely SM local group $\cG_{SM}$, the low-energy expression for $\UH(x)$ introduces a scale $v$ and not $\fL$ for the eaten GBs, that appears through a redefinition of the GB fields in order to have canonically normalized kinetic terms. In here it will be kept generically the scales $\fLR$ as the associated GBs will be canonically normalized when writing their corresponding kinetic terms later on.} as it is customary via the dimensionless unitary matrix $\UH(x)$, more specifically through $\UHL(x)$ and $\UHR(x)$ for the symmetry group $SU(2)_L\otimes SU(2)_R$, and defined as
\beq
\UHLR\,(x)=e^{i\,\tau_a\,\pi^a_{L(R)}(x)/\fLR}\, , 
\label{Goldstone-matrices}
\eeq

\nt with $\tau^a$ Pauli matrices and $\pi^a_{L(R)}(x)$ the corresponding Goldstone bosons fields suppressed by their associated non--linear sigma model scale $\fLR$. In this set--up, the scale $f_R$ comes from the additional Goldstone boson dynamics introduced by $SU(2)_R$ group. The aforementioned local symmetry for this set-up, i.e., the local group $\cG=SU(2)_L\otimes SU(2)_R\otimes U(1)_{B-L}$, will introduce local rotations as 
\be
\LT(x)\equiv e^{\frac{i}{2}\tau^a\alpha^a_L(x)},\, 
\RT(x)\equiv e^{\frac{i}{2}\tau^a\alpha^a_R(x)},\, 
\UY(x)\equiv e^{\frac{i}{2}\tau^3\alpha^0(x)}
\label{Local-transformations}
\ee
\nt with $\alpha^a_{L,R}(x)$ and $\alpha^0(x)$ space-time dependent variables parametrizing the local symmetry transformations, and leading thus to the Goldstone boson matrices $\UH_{L(R)}(x)$ to transform accordingly as
\be
\UH_L \rightarrow \mathbf{L}\,\UH_L\,\mathbf{U}^\dagger_Y\,, \qquad\qquad
\UH_R \rightarrow \mathbf{R}\,\UH_R\,\mathbf{U}^\dagger_Y
\label{U-transformation-properties}
\ee

\nt where hereafter the local space-time dependence of the implied objects will be omitted. The adimensionality of the $SU(2)_{L(R)}$ Goldstone matrices $\UHLR(x)$ is the technical key to understand why the dimension of the leading low-energy operators describing the dynamics of the scalar 
sector differs for a non--linear Higgs sector~\cite{Appelquist:1980vg,Longhitano:1980iz,Longhitano:1980tm,
Feruglio:1992wf,Appelquist:1993ka} and a purely linear regime. In the former, non-renormalizable operators 
containing extra powers of a light $h$ are weighted by powers of $h/\fL$~\cite{Georgi:1984af}, while 
the Goldstone boson contributions encoded in $\UHLR(x)$ do not exhibit any scale suppression. In the linear regime, instead, the light $h$ and the three SM GBs are encoded into the scalar doublet $H$, with mass dimension one: therefore any extra insertion of $H$ is suppressed by a power of the NP cutoff scale.

For the non--linear EFT construction undertaken in here, it is natural at this point to introduce the corresponding covariant derivative objects for both of the Goldstone matrices $\UHLR(x)$ as
\be
\begin{aligned}
&\DL^\mu \UHLR \equiv \\
&\derp^\mu \UHLR \, + \,\frac{i}{2}\,\gLR\,W^{\mu,a}_{L(R)}\,\tau^a\,\UHLR - \frac{i}{2}\,g'\,B^\mu\, \UHLR\,\tau^3 \,.
\end{aligned}
\label{Covariant-derivatives}
\ee

\nt with the $SU(2)_L$, $SU(2)_R$ and $U(1)_{B-L}$ gauge fields denoted by $W^{a\mu}_L$, $W^{a\mu}_R$ and $B^\mu$ correspondingly, and the associated gauge couplings $\gL$, $\gR$ and $g'$ respectively. Aimed by the definitions in Eq.~\eqref{Covariant-derivatives}, it is straightforward to introduce in the framework the commonly used covariant quantities $\VL$ and $\TL$, now enlarged to cover the adjoints $SU(2)_{L(R)}$--covariant vectorial $\VLLRmuu$ and 
the covariant scalar $\TLLR$ objects as
\be
\VLLRmuu \equiv \left(\DL^\mu \UHLR\right)\,\UHLR^\dagger\,, \quad
\TLLR \equiv \UHLR\,\tau_3\,\UHLR^\dagger\,,
\label{EFT-building-blocks}
\ee

\nt transforming correspondingly under the local 
$\cG$ transformation of Eq.~\eqref{Local-transformations} as

\be
\begin{aligned}
& \VLLmuu \rightarrow \quad\LT\,\VLLmuu\,\LT^\dagger\,,\qquad 
&& \VLRmuu \rightarrow \quad\RT\,\VLRmuu\,\RT^\dagger\,, \\[1mm] 
& \TLL \rightarrow \quad\LT\,\TLL\,\LT^\dagger\,,\qquad 
&& \TLR \rightarrow \quad\RT\,\TLR\,\RT^\dagger\,.
\end{aligned}
\label{V-T-transformation-properties}
\ee

\nt The covariant objects described in Eq.~\eqref{EFT-building-blocks}, constitute the basic building blocks for the non--linear EFT construction considered in here. 

Notice that the local gauge invariance of the theory allows to build operators made out of traces depending either on products of purely left-handed or right-handed covariant objects. As soon as the operators mixing the left and right-handed structures are considered, new covariant objects emerge to fully guarantee the constructions of such mixed operators. A feature that will be described in the following.


\subsection{$SU(2)_L-SU(2)_R$ interplay}
\label{Interplay}

\nt Consider a set of left and right--handed covariant objects $\cOL^i$ and $\cOR^i$ (with $i$ labelling either a scalar, vector, or tensor object) transforming under the local symmetries of the model as in Eq.~\eqref{V-T-transformation-properties}. Traces like $\Tr\left(\cOLR^i\,\cOLR^j\,...\,\cOLR^n\right)$ or the product of traces $\Tr\left(\cOL^i\,\cOL^j\,...\,\cOL^n\right)\Tr\left(\cOR^k\,\cOR^l\,...\,\cOR^m\right)$
are manifestly gauge invariant under $\cG$, and will be useful when building up the complete tower of non--linear effective operators. Nonetheless, as soon as the mixed structures are considered, e.g. the simple trace $\Tr\left(\cOL^i\,\cOR^j\right)$, the gauge invariance is evidently no longer satisfied. To recover it back at the level of these type of traces, proper insertions of the Goldstone matrices $\UHL$ and $\UHR$ are in order as 
\be
\Tr\left(\cOL^i\,\cOR^j\right) \quad\rightarrow \quad\Tr\left(\UHL^\dagger\,\cOL^i\,\UHL\,\UHR^\dagger\,\cOR^j\,\UHR\right)
\label{Mixed-structures}
\ee

\nt the final trace is now manifestly permitted by the local symmetries of the theory and it is well behaved under such transformations. Proper insertions of $\UHL$ and $\UHR$ as in Eq.~\eqref{Mixed-structures}, lead directly to introduce the following objects
\be
\cOLt^i \equiv \UHL^\dagger\,\cOL^i\,\UHL\,, \qquad\qquad
\cORt^i \equiv \UHR^\dagger\,\cOR^i\,\UHR\,,
\label{U-transforming-objects}
\ee

\nt that are in order henceforth for the construction of operators made out of mixed $SU(2)_L$ and $SU(2)_R$ covariant structures
, i.e. made out of mixed products of $\cOL^i$ and $\cOR^j$. Notice that the new defined objects are transforming under the local $\cG$--transformations of Eq.~\eqref{Local-transformations} as
\be
\cOLt^i \rightarrow \mathbf{U}_Y\,\cOLt^i\,\mathbf{U}^\dagger_Y\,, \qquad\qquad
\cORt^i \rightarrow \mathbf{U}_Y\,\cORt^i\,\mathbf{U}^\dagger_Y
\label{Utilde-transformation-properties}
\ee

\nt This feature allows to bring more invariant objects to play a role into the game. In fact, following the definitions in Eq.~\eqref{U-transforming-objects} for the covariant vectorial $\VLLRmuu$ and scalar $\TLLR$ objects already defined in Eq.~\eqref{EFT-building-blocks}, the new quantities are found
\be
\begin{aligned}
\VLLRmuut &\equiv \UHLR^\dagger\,\VLLRmuu\,\UHLR = -\left(\DL^\mu \UHLR \right)^\dagger\UHLR
\end{aligned}
\label{Vtilde}
\ee

\nt and 
\be
\TLLRt \equiv \UHLR^\dagger\,\TLLR\,\UHLR = \tau_3\,,
\label{Ttilde}
\ee

\nt where the unitary property of the Goldstone matrices $\UHLR$  has been employed in addition.

As conclusion from this section, it is possible to infer therefore the mandatory introduction of these extra covariant objects $\cOLRt^i$ in order to construct any possible operator mixing the left and right-handed covariant quantities $\cOLR^i$. Notice that under the new definitions in Eq.~\eqref{U-transforming-objects}, traces like $\Tr\left(\cOLRt^i\,\cOLRt^j\,...\,\cOLRt^n\right)$ or the product of traces $\Tr\left(\cOLt^i\,\cOLt^j\,...\,\cOLt^n\right)\Tr\left(\cORt^k\,\cORt^l\,...\,\cORt^m\right)$, will turn out to be equivalent to $\Tr\left(\cOLR^i\,\cOLR^j\,...\,\cOLR^n\right)$ and $\Tr\left(\cOL^i\,\cOL^j\,...\,\cOL^n\right)\Tr\left(\cOR^k\,\cOR^l\,...\,\cOR^m\right)$ respectively. Therefore, operators made out of products of purely left or right-handed covariant quantities can be constructed out also via the covariant objects $\cOLRt^i$ defined in Eq.~\eqref{U-transforming-objects}. Specifically for the framework assumed in here, the EFT building blocks can be established through Eqs.~\eqref{Vtilde}-\eqref{Ttilde} 
\be
\VLLRmuut\,, \quad  \TLLRt \,,
\label{Larger-EFT-building-blocks}
\ee
\nt and transforming correspondingly as

\be
\begin{aligned}
& \VLLmuut \rightarrow \quad \mathbf{U}_Y\,\VLLmuut\,\mathbf{U}^\dagger_Y\,,\qquad 
&& \VLRmuut \rightarrow \quad \mathbf{U}_Y\,\VLRmuut\,\mathbf{U}^\dagger_Y\,, \\[1mm] 
& \TLLt \rightarrow \quad \mathbf{U}^\dagger_Y
\,\TLLt\,\mathbf{U}^\dagger_Y
\,,\qquad 
&& \TLRt \rightarrow \quad \mathbf{U}_Y
\,\TLRt\,\mathbf{U}^\dagger_Y
\,.
\end{aligned}
\label{New-V-T-transformation-properties}
\ee

\nt Similar definitions for the strength gauge fields $\WWLRu$ are straightforward
\be
\WWLRut \equiv \UHLR^\dagger\,\WWLRu\,\UHLR\,.
\label{Wtilde}
\ee

\nt Henceforth the set given in Eq.~\eqref{Larger-EFT-building-blocks} and Eq.~\eqref{Wtilde} will shape the building blocks for the construction of the effective EW non--linear left-right symmetric approach undertaken in this work\footnote{Definitions in Eqs.~\eqref{Vtilde}-\eqref{Ttilde} and Eq.~\eqref{Wtilde}, are basically those ones employed from the very beginning in the non--linear effective treatment of Refs.~\cite{Zhang:2007xy,Wang:2008nk}. In fact, $\VLLRmuut$ in Eq.~\eqref{Vtilde} turns out to be equivalent to the element $X^\mu_i$, whereas $\WWLRut$ are proportional to $\overline{W}^{\mu\nu}_i$ ($i=L,R$), introduced both in~\cite{Zhang:2007xy,Wang:2008nk}. This point will be remarked later again when listing all the non--linear operators mixing the left and right-handed covariant objects up to the $p^4$--order in the effective Lagrangian expansion.}. Construction that will be enlarged after accounting for all the possible gauge-Higgs couplings arising out in this scenario through the generic polynomial light Higgs function $\cF(h)$~\cite{Alonso:2012px}, singlet under $\cG$, introduced in the following section and later on when listing the non--linear left-right symmetric operators up to the $p^4$-order in the Lagrangian expansion.


\subsection{Light Higgs inclusion}
\label{Light-Higgs}

\nt Higgs couplings can be directly brought into the scenario via the generic light Higgs polynomial functions $\cF(h)$~\cite{Alonso:2012px} expandable as customary
\beq
\cF_i(h)\equiv1+2\,{a}_i\,\frac{h}{\fL}+{b}_i\,\frac{h^2}{\fL^2}+\ldots\,,
\label{F}
\eeq
\nt with dots standing for higher powers in $h/\fL$~\cite{Georgi:1984af} that will not be considered below. The scale suppression for each $h$--insertion is  dictated by $\fL$, as it is the scale where $h$ is generated as a GB prior to the extension of the SM local group $\cG_{SM}$ to the larger one $\cG$. Gauge--$h$ interactions will arise by letting the non--linear operators to be coupled either directly to $\cF(h)$ or through its derivative couplings, e.g. via $\partial^\mu \cF(h)$, $\left(\partial^\mu \cF(h)\right)^2$, $\partial^\mu \cF(h)\,\partial_\mu \cF(h)$ and $\partial^\mu\partial_\mu \cF(h)$. Thus the building blocks set in Eqs.~\eqref{Larger-EFT-building-blocks} and~\eqref{Wtilde} gets complemented by accounting for all the possible interactions from $\cF(h)$  and its corresponding derivative couplings, under the assumption of a CP--even behaviour for $h$.

So far the local gauge symmetry $\cG$ has been demanded throughout the non--linear effective approach considered in here. There is another relevant symmetry emerging after enlarging the initial SM local group $\cG_{SM}$ to $\cG$, precisely the one related with the exchange of the left and right--handed components of the group $SU(2)_L\otimes SU(2)_R$: the parity symmetry $P_{LR}$ introduced in the following.


\subsection{$P_{LR}$--symmetry}
\label{P-LR-symmetry}

\nt The discrete parity symmetry $P_{LR}$ will imply the exchange of the $L$ and $R$ components of $SU(2)_L\otimes SU(2)_R$ and will play as well an important role for protecting the $Zb\bar b$--coupling from large corrections in the context of composite Higgs models~\cite{Agashe:2006at}. Moreover, it has been shown to be an accidental symmetry up to $p^2$--Goldstone Lagrangian, broken by several $p^4$--operators (even in the limit $\gL$, $\gR$, $g'\rightarrow 0$), with leading $p^6$ non-zero effects in $\pi\pi$ scattering in the context of a general effective $SO(5)/SO(4)$ composite Higgs model scenario~\cite{Contino:2011np}. Expectedly, the effective approach proposed in this work will share these features as $SU(2)_L \otimes SU(2)_R \sim SO(4)$. In fact, before gauging the scenario, i.e for the case of vanishing couplings $\gL$, $\gR$ and $g'$, the action of the discrete parity symmetry $P_{LR}$ will lead to the traces like $\Tr\left(\cOL^i\,\cOL^j\,...\,\cOL^n\right)$ to be mapped onto $\Tr\left(\cOR^i\,\cOR^j\,...\,\cOR^n\right)$ and vice versa, or the product of traces $\Tr\left(\cOL^i\,\cOL^j\,...\,\cOL^n\right)\Tr\left(\cOR^k\,\cOR^l\,...\,\cOR^m\right)$ to be $P_{LR}$-even. Furthermore, it will lead to the traces made out of the mixed product of left and right-handed objects, e.g. $\Tr\left(\cOLt^i\,\cORt^j\right)$ to give rise to the trace\footnote{Superscripts $i$ and $j$ involve Lorentz indices, so their naive exchange does not recover the initial trace necessarily.} $\Tr\left(\cOLt^j\,\cORt^i\right)$ and therefore more possible effective operators are feasible as it will be realized when listing the operators afterwards.

The EW effective Lagrangian described in the following section keeps track of the light dynamical Higgs picture in~\cite{Alonso:2012px,Brivio:2013pma,Gavela:2014vra,Alonso:2012pz} (see also Ref.~\cite{Buchalla:2013rka,Buchalla:2013eza}, and for a short summary on the subject~\cite{Brivio:2015hua}), focused only in the CP--conserving bosonic operators\footnote{Look at~\cite{Cvetic:1988ey,Alonso:2012jc,Alonso:2012pz,Buchalla:2012qq,
Buchalla:2013rka} for non--linear analysis including fermions.}, including the new spin--1 resonances sourced by the symmetry group $SU(2)_R$, and up to $p^4$--order in the Lagrangian expansion.


\section{The effective Lagrangian}
\label{EffectiveLagrangian}

\nt Effective NP contributions from the strong dynamics assumed in here will lead to non--zero departures with respect to the SM Lagrangian $\LL_0$ and will be encoded in the Lagrangian $\LL_\text{chiral}$ through
\beq
\LL_\text{chiral} = \LL_0\,+\,\LL_{0,R}
\,+\,\LL_{0,LR}
\,+\,\Delta \LL_{\text{CP}}\,+\,\Delta\LL_{\text{CP},LR}\,.
\label{Lchiral}
\eeq 

\nt As for the bosonic interacting sector concerns, the first piece in $\LL_\text{chiral}$ reads as
\be
\begin{aligned}
\LL_0 &= -\dfrac{1}{4}\,\BBd\,\BBu-\dfrac{1}{4}\,W^a_{\mu\nu,\,L}\,W^{\mu\nu,\,a}_L-
\dfrac{1}{4}\,G^a_{\mu\nu}\,G^{\mu\nu,\,a}\,+\\[2mm]
&+\frac{1}{2} (\derp_\mu h)(\derp^\mu h) - V (h)-\dfrac{\fL^2}{4}\Tr\Big(\VLLmuu\VLLmud\Big)\left(1+\frac{h}{\fL}\right)^2
\end{aligned}
\label{LLO}
\ee

\nt providing the SM strength gauge kinetic terms canonically normalized in the first line, whereas $h$--kinetic terms and the effective scalar potential $V(h)$ triggering the EWSB from the first two terms at the second line, plus the $W^\pm_L$ and $Z_L$ masses (before considering the corresponding right handed and mixed left--right handed terms introduced a posteriori) and their couplings to $h$ from the last term in the second line of Eq.~\eqref{LLO}. To keep clear the notation according to the assumed local symmetry group $\cG$, the $SU(2)_L$--kinetic term and the custodial conserving $p^2$--operator $\Tr\left(\VL^\mu\,\VL_\mu\right)$ have been properly labelled, but the Lagrangian $\LL_0$  as to emphasize its LO character once the purely SM local group $\cG_{SM}$ is considered singly. From the scale factor of $\Tr\left(\VLLmuu\,\VLLmud\right)$ it is inferred that GB--kinetic terms are already canonically normalized, in agreement with $\UHL$--definition of Eq.~\eqref{Goldstone-matrices}.

Phenomenological bounds strongly constrain the custodial breaking $p^2$--operator, being thus left for the NP departures analysed later~\cite{Brivio:2013pma}. As long as the scenario calls for the symmetric counterpart sourced by the corresponding local $SU(2)_R$--extension, non--zero NP departures with  respect to those from $\LL_0$ will play a role into the game, being encoded through the remaining pieces of $\LL_\text{chiral}$ Eq.~\eqref{Lchiral}, focused only on the CP--conserving operators, and defined by 

\begin{itemize}

\item $\LL_{0,R}$: accounting for all the non--linear $p^2$-operators dictated by the $SU(2)_R$--extension,

\item $\LL_{0,LR}$: describing up to $p^2$--order all the operators mixing the left and right handed-covariant objects,

\item $\Delta\LL_{\text{CP}}$: encoding all those effective non--linear operators made out of purely left or right handed covariant objects up to the $p^4$-order, and 

\item $\Delta\LL_{\text{CP},LR}$: parametrizing any possible mixing interacting term between the $SU(2)_L$ and $SU(2)_R$-covariant objects up to $p^4$-operators and permitted by the underlying left-right symmetry. 

\end{itemize}

\nt The following sections describe all these components one per one.


\boldmath
\subsection{$SU(2)_R$--extension of $\LL_0$: $\LL_{0,R}$}
\unboldmath

\nt The LO $p^2$--Lagrangian for the $SU(2)_R$--extension of the framework is going to be encoded by  $\LL_{0,R}$, and described as 
\beq
\LL_{0,R}= -\dfrac{1}{4}\,W^a_{\mu\nu,\,R}\,W^{\mu\nu,\,a}_R\,-\,\frac{\fR^2}{4}\,\Tr\Big(\VLRmuu\,\VLRmud\Big)\left(1+\frac{h}{\fL}\right)^2
\label{LLO-Right}
\eeq
\nt where both of the terms in Eq.~\eqref{LLO-Right} are accounting for the corresponding $SU(2)_R$--counterparts for the $SU(2)_L$ strength gauge kinetic term and the custodial conserving operator $\Tr\Big(\VLLmuu\,\VLLmud\Big)$ from the LO $p^2$--Lagrangian $\LL_0$ of Eq.~\eqref{LLO}. Notice that the second operator in $\LL_{0,R}$, i.e. the custodial $p^2$--operator $\Tr\Big(\VLRmuu\,\VLRmud\Big)$ sourced by the $\cG$-invariance of the approach, entails an additional scale $\fR$ that encodes the new high energy scale effects introduced in the scenario once the SM local symmetry group $\cGSM$ is extended to $\cG$, and will provide canonically normalized right handed GB according to the $\UHR$-definition in Eq.~\eqref{Goldstone-matrices}. Moreover, the way the right handed gauge fields couple to the light Higgs $h$ is dictated by following its left handed counterpart in $\LL_0$ Eq~\eqref{LLO}.


\boldmath
\subsection{$SU(2)_L-SU(2)_R$ $p^2$--interplay: $\LL_{0,LR}$}
\unboldmath

\nt Up to the $p^2$--order expansion, operators made of mixed products of left and right--handed objects emerge from the covariant objects $\cOLRt$ defined in Eq.~\eqref{U-transforming-objects}, specifically from $\VLLRmuut$, $\TLLRt$ and $\WWLRut$ defined in Eqs.~\eqref{Vtilde}, \eqref{Ttilde} and \eqref{Wtilde} respectively. In fact, the $p^2$--interplaying Lagrangian for such contributions is parametrized via
\beq
\begin{aligned}
&\LL_{0,LR}=\\ 
&-\dfrac{1}{2}\,\Tr\left(\WWLut\,\WWRdt\right)-\frac{\fL\fR}{2}\,\Tr\Big(\VLLmuut\VLRmudt\Big)\left(1+\frac{h}{\fL}\right)^2
\end{aligned}
\label{LLO-Left-Right}
\eeq
\nt with $\WWchiut\equiv \widetilde{W}^{\mu\nu,a}_\chi\tau^a/2$ ($\chi$ standing for $\chi=L,R$),  and  where the trace has explicitly been written in the first term as to make clear its $\cG$-invariance under the property transformations in Eq.~\eqref{Utilde-transformation-properties}. Such term is not present among the $p^2$--terms listed in Refs.~\cite{Zhang:2007xy,Wang:2008nk}, and will be explicitly kept hereafter in this work. The gauge-$h$ interactions in the second term are following the same prescription in analogy of $\LL_0$ and $\LL_{0,R}$. As it can be clearly seen, in the unitary gauge the $p^2$--Lagrangian $\LL_{0,LR}$ drives non-zero kinetic mixing terms between the strength gauge fields $\WWLu$ and $\WWRu$ from the first term in Eq.~\eqref{LLO-Left-Right}, leading thus to rotate the strength gauge sector in order to make it diagonal. Likewise, non-zero mass mixing terms among the left and right--handed gauge fields are triggered by the second term in $\LL_{0,LR}$, entailing thus an additional diagonalization in the gauge sector in order to obtain the required physical gauge masses. Such rotation and diagonalization issues are left for a future work~\cite{Shu:2015cxm}, as the aim at this work consists in establishing the whole $\cG$--invariant non--linear operator basis up to the $p^4$--contributions for the EW left--right bosonic interactions faced here.
 
At higher orders in the momentum expansion more interactions are sourced by the local symmetry group $\cG$, part of them accounted by the fourth piece of $\LL_\text{chiral}$ in Eq.~\eqref{Lchiral}, i.e.  $\Delta\LL_{\text{CP}}$, and described in the following.


\boldmath
\subsection{$\cG$--extension of $\LL_0\,+\,\LL_{0,R}\,+\,\LL_{0,LR}$: $\Delta\LL_{\text{CP}}$}
\unboldmath

\nt  $\Delta\LL_{\text{CP}}$ includes all the possible bosonic CP--conserving non--linear operators (that is, all pure gauge and gauge--$h$ operators plus pure Higgs ones) that describe deviations from the LO $p^2$--picture $\LL_0\,+\,\LL_{0,R}\,+\,\LL_{0,LR}$, due to the strong interacting physics present at scales higher than the EW one, and allowed by the $\cG$--invariance of the framework. All the possible CP--conserving gauge--$h$ interactions up to $p^4$--operators are split in here as
\be
\Delta \LL_{\text{CP}}=\Delta \LL_{\text{CP},L}+\Delta \LL_{\text{CP},R}
\label{DeltaL-CP-even}
\ee
\nt where the suffix $L(R)$ labels all those operators set constructed out by means of the $SU(2)_{L(R)}$ building blocks in Eqs.~\eqref{Larger-EFT-building-blocks} and~\eqref{Wtilde}. In the context of purely EW chiral effective theories coupled to a light Higgs, the first contribution to $\Delta \LL_{\text{CP}}$, i.e. $\Delta \LL_{\text{CP},L}$, has already been provided in Refs.~\cite{Alonso:2012px,Brivio:2013pma}, whereas for the left--right symmetric frameworks, part of $\Delta \LL_{\text{CP},L}$ and $\Delta \LL_{\text{CP},R}$ were already analysed in Refs.~\cite{Zhang:2007xy,Wang:2008nk} as it will be shown later. Both of the contributions $\Delta \LL_{\text{CP},L}$ and $\Delta \LL_{\text{CP},R}$, parametrizing the NP deviations with respect to the LO $p^2$-Lagrangian $\LL_0\,+\,\LL_{0,R}\,+\,\LL_{0,LR}$, can be written down as 
\beq
\begin{aligned}
\Delta \LL_{\text{CP},L}&=c_G\cP_G(h)+c_B\cP_B(h)+\sum_{i=\{W,C,T\}}c_{i,L}\cP_{i,L}(h)\,\,+ \\[2mm]
&+\sum_{i=1}^{26} c_{i,L}\cP_{i,L}(h) \,+\, c_H \cP_H(h) \,+\, c_{\Box H} \cP_{\Box H}(h)
\label{DeltaL-CP-even-L}
\end{aligned}
\eeq

\nt and

\be
\Delta\LL_{\text{CP},R}= \sum_{i=\{W,C,T\}}c_{i,R}\cP_{i,R}(h)\,\, \,+\,\,\,\sum_{i=1}^{26} c_{i,R}\cP_{i,R}(h)
\label{DeltaL-CP-even-R}
\ee

\nt where $c_{B}$, $c_{G}$ and  $c_{i,\chi}$ are model--dependent constant coefficients, whilst the first line of $\Delta \LL_{\text{CP},L}$ in Eq.~\eqref{DeltaL-CP-even-L} and the first term in Eq.~\eqref{DeltaL-CP-even-R} can be jointly written as
\beq
\begin{aligned}
\cP_G(h)\,\,  &= -\frac{g^2_s}{4}\,G_{\mu\nu}^a\,G^{\mu\nu}_a\,\cF_G(h) \\
\cP_B(h)\,\,  &= -\frac{g'^2}{4}\,B_{\mu\nu}\,B^{\mu\nu}\,\cF_B(h) \\ 
\cP_{W,\,\chi}(h)\,\,  &= -\frac{\gchi^2}{4}\, W_{\mu\nu,\,\chi}^a\,W^{\mu\nu,\,a}_\chi\,\cF_{W,\,\chi}(h)  \\  
\cP_{C,\,\chi}(h)\,\,  &= - \frac{\fchi^2}{4}\Tr\Big(\VLchimuu\,\VLchimud\Big) \,\cF_{C,\,\chi}(h) \\ 
\cP_{T,\,\chi}(h)\,\,  &= \frac{\fchi^2}{4}\, \Big(\Tr\Big(\TLchi\,\VLchimuu\Big)\Big)^2\,\cF_{T,\,\chi}(h)  
\label{GT}
\end{aligned}
\eeq
\nt with suffix $\chi$ labelling again as $\chi=L,R$, and the generic $\cF_i(h)$--function of the scalar singlet $h$ is defined for all the operators following Eq.~\eqref{F}. Notice that operators in Eq.~\eqref{GT} account for the CP conserving kinetic gauge terms $\{\cP_G(h),\,\cP_B(h),\,\cP_{W,\,\chi}(h)\}$, custodial conserving $\cP_{C,\,\chi}(h)$  and custodial breaking operators $\cP_{T,\,\chi}(h)$, all of them coupled to their corresponding $\cF_i(h)$. 

The complete linearly independent set of 26 CP--conserving pure gauge and gauge--$h$ non--linear $\cG$--invariant operators up to the $p^4$-order in the effective Lagrangian expansion, and encoded by $\cP_{i,\,L}(h)$ (first term in the second line of $\Delta \LL_{\text{CP},\,L}$, Eq.~\eqref{DeltaL-CP-even-L}) have completely been listed in  Refs.~\cite{Alonso:2012px,Brivio:2013pma}. On the other hand, the symmetric counterpart extending the aforementioned set $\cP_{i,\,L}(h)$ and accounting for all the possible CP--conserving pure gauge and gauge--$h$ interactions up to the $p^4$--operators made out of the corresponding $SU(2)_R$--building blocks in Eqs.~\eqref{Larger-EFT-building-blocks} and~\eqref{Wtilde}, is described by the complete linearly independent set of 26 CP-conserving operators $\cP_{i,\,R}(h)$ (second term in $\Delta \LL_{\text{CP},\,R}$ of Eq.~\eqref{DeltaL-CP-even-R}), so all in all there are 52 non--linear operators in total, among the which 38 of them (19 $\cP_{i,\,L}(h)$ + 19 $\cP_{i,\,R}(h)$) had already been listed in  Refs.~\cite{Zhang:2007xy,Wang:2008nk}. In here, 14 additional operators have been found (7 $\cP_{i,\,L}(h)$ + 7 $\cP_{i,\,R}(h)$) with respect to Refs.~\cite{Zhang:2007xy,Wang:2008nk} (among the which the seven operators corresponding to $\chi = L$ were already reported in Refs.~\cite{Alonso:2012px,Brivio:2013pma}) 
and naturally promoted by the symmetries of the model (together with the $P_{LR}$--symmetry), such that the whole tower of operators making up the basis $\{\cP_{i,\,L}(h),\,\cP_{i,\,R}(h)\}$ is given by:
\begin{widetext}
\beq
\hspace{-0.5cm}
\begin{aligned}
&\cP_{1,\,\chi}(h) = \gchi\,g' \,B_{\mu\nu}\,\Tr\Big(\TLchi\,\WWchiu\Big)\,\cF_{1,\,\chi}(h)\,,
&&\cP_{14,\,\chi}(h)=\gchi\,\epsilon_{\mu\nu\rho\sigma}\,\Tr\Big(\TLchi\,\VLchimuu\Big)\,\Tr\Big(\VLchinuu\,\WWWchiu\Big)\,\cF_{14,\,\chi}(h)\,, \\[1.4mm] 
&\cP_{2,\,\chi}(h) = i\,g' \,B_{\mu\nu}\,\Tr\Big(\TLchi\,\Big[\VLchimuu,\VLchinuu\Big]\Big)\,\cF_{2,\,\chi}(h)\,,
&&\brown{\cP_{15,\,\chi}(h) = \Big(\Tr\Big(\TLchi\,\cD_\mu\VLchimuu\Big)\Big)^2\,\cF_{15,\,\chi}(h)}\,, \\[1.4mm] 
&\cP_{3,\,\chi}(h)  = i\,\gchi\,\Tr\Big(\WWchiu\,\Big[\VLchimud,\VLchinud\Big]\Big)\,\cF_{3,\,\chi}(h)\,,
&&\brown{\cP_{16,\,\chi}(h) = \Tr\Big(\Big[\TLchi,\VLchinud\Big]\,\cD_\mu \VLchimuu\Big) \, \Tr\Big(\TLchi\,\VLchinuu\Big)\,\cF_{16,\,\chi}(h)}\,,\\[1.4mm]
&\cP_{4}(h)= i\,g'\,B_{\mu\nu}\,\Tr\Big(\TLchi\,\VLchimuu\Big)\,\derp^\nu \cF_{4}(h)\,,
&&\cP_{17,\,\chi}(h) = i\,\gchi \,\Tr\Big(\TLchi\,\WWchiu\Big)\,\Tr\Big(\TLchi\,\VLchimud\Big)\,\derp_\nu \cF_{17,\,\chi}(h)\,, \\[1.4mm]
&\cP_{5,\,\chi}(h) = i\,\gchi \,\Tr\Big(\WWchiu\,\VLchimud\Big)\,\derp_\nu \cF_{5,\,\chi}(h)\,,
&&\cP_{18,\,\chi}(h) = \tr\Big(\TLchi\Big[\VLchimuu,\VLchinuu\Big]\Big)\,\tr\Big(\TLchi\,\VLchimud\Big)\, \derp_\nu\cF_{18,\,\chi}(h)\,,  \\[1.4mm] 
&\cP_{6,\,\chi}(h) =\Big(\Tr\Big(\VLchimud\,\VLchimuu\Big)\Big)^2\,\cF_{6,\,\chi}(h)\,,
&&\brown{\cP_{19,\,\chi}(h) = \Tr\Big(\TLchi\,\cD_\mu\VLchimuu\Big)\,\Tr\Big(\TLchi\,\VLchinuu\Big)\,\derp_\nu \cF_{19,\,\chi}(h)}\,, \\[1.4mm] 
&\brown{\cP_{7,\,\chi}(h) =\Tr\Big(\VLchimud\,\VLchimuu\Big)\,\derp_\nu\derp^\nu\cF_{7,\,\chi}(h)}\,,
&&\cP_{20,\,\chi}(h) = \tr\Big(\VLchimud\,\VLchimuu\Big)\,\derp_\nu\cF_{20,\,\chi}(h)\derp^\nu\cF_{20,\,\chi}'(h)\,, \\[1.4mm]
&\cP_{8,\,\chi}(h) =\Tr\Big(\VLchimuu\,\VLchinuu\Big)\,\derp_\mu\cF_{8,\,\chi}(h)\,\derp_\nu\cF'_{8,\,\chi}(h)\,,
&&\cP_{21,\,\chi}(h) = \Big(\tr\Big(\TLchi\,\VLchimuu\Big)\Big)^2\,\derp_\nu\cF_{21,\,\chi}(h)\,\derp^\nu\cF_{21}'(h)\,, \\[1.4mm] 
&\brown{\cP_{9,\,\chi}(h) = \Tr\Big(\Big(\cD_\mu\VLchimuu\Big)^2 \Big)\,\cF_{9,\,\chi}(h)} \,,
&&\cP_{22,\,\chi}(h) = \Big(\tr\Big(\TLchi\,\VLchimuu\Big)\,\derp_\mu\cF_{22,\,\chi}(h)\Big)^2\,,\\[1.4mm] 
&\brown{\cP_{10,\,\chi}(h) =\Tr\Big(\VLchinuu\,\cD_\mu\VLchimuu\Big)\,\derp_\nu \cF_{10,\,\chi}(h)}\,,
&&\cP_{23,\,\chi}(h) = \tr\Big(\VLchimud\,\VLchimuu\Big)\,\Big(\tr\Big(\TLchi\VLchinuu\Big)\Big)^2\, \cF_{23,\,\chi}(h)\,,\\[1.4mm] 
&\cP_{11,\,\chi}(h) = \Big(\tr\Big(\VLchimuu\,\VLchinuu\Big)\Big)^2\,\cF_{11,\,\chi}(h)\,,
&&\cP_{24,\,\chi}(h) = \tr\Big(\VLchimuu\,\VLchinuu\Big)\,\tr\Big(\TLchi\,\VLchimud\Big)\,\tr\Big(\TLchi\,\VLchinud\Big)\,\cF_{24,\,\chi}(h)\,,\\[1.4mm] 
&\cP_{12,\,\chi}(h) = \gchi^2\,\Big(\Tr\Big(\TLchi\,\WWchiu\Big)\Big)^2\,\cF_{12,\,\chi}(h)\,,
&&\brown{\cP_{25,\,\chi}(h)= \Big(\tr\Big(\TLchi\,\VLchimuu\Big)\Big)^2\,\derp_\nu\derp^\nu\cF_{25,\,\chi}(h)}\,,\\[1.4mm] 
&\cP_{13,\,\chi}(h) = i\,\gchi\,\Tr\Big(\TLchi\,\WWchiu\Big)\,\Tr\Big(\TLchi\,\Big[\VLchimud,\VLchinud\Big]\Big)\,\cF_{13,\,\chi}(h)\,,
&&\cP_{26,\,\chi}(h)=\Big(\Tr\Big(\TLchi\,\VLchimuu\Big)\,\Tr\Big(\TLchi\,\VLchinuu\Big)\Big)^2\,\cF_{26,\,\chi}(h)\,, 
\end{aligned}
\label{CP-even-basis}
\eeq
\end{widetext}
\nt with $\WWchiu\equiv W^{\mu\nu,a}_\chi\tau^a/2$. In red color have been highlighted all those operators already listed in the context of purely EW chiral effective theories coupled to a light Higgs in Refs.~\cite{Alonso:2012px,Brivio:2013pma} (for $\chi=L$) and not provided in the left-right symmetric EW chiral treatment of Refs.~\cite{Zhang:2007xy,Wang:2008nk}. In Eq.~\eqref{CP-even-basis}, $\DLL_\mu$ denotes the covariant derivative on a field transforming 
in the adjoint representation of $SU(2)_L$, and defined as
\be
\DLL^\mu \VLchinuu \equiv \partial^\mu \VLchinuu\,+\,i\,\gchi\left[ W^\mu_\chi, \VLchinuu \right]\,,\quad \chi=L,R\,.
\label{DV-covariant-derivative}
\ee

Notice that operators $\{\cP_{9,\,\chi}(h),\,\cP_{15-16,\,\chi}(h)\,\cP_{19,\,\chi}(h)\}$ containing the contraction $\cD_\mu \VLchimuu$, and $\{\cP_{7,\,\chi}(h),\,\cP_{25,\,\chi}(h)\}$  with double derivatives of $\cF(h)$, are not present in  Refs.~\cite{Zhang:2007xy,Wang:2008nk}, and are the resulting additional ones from the $SU(2)_R$-extension of $\Delta\LL_{\text{CP},\,L}$, being naturally allowed by the local symmetries of the model. Notice as well that the entire basis $\cP_{i,\,R}(h)$ contained in Eq.~\eqref{CP-even-basis} (for $\chi=R$) comes out just from the straightforward parity action under $P_{LR}$ of the operators tower $\cP_{i,\,L}$ in Refs.~\cite{Alonso:2012px,Brivio:2013pma,Gavela:2014vra}, or in other words, the whole basis $\cP_{i,\,R}(h)$ is mapped from $\cP_{i,\,L}(h)$ via $P_{LR}$--transformation.

Finally, the last two terms in the second line of $\Delta \LL_{\text{CP},L}$ in Eq.~\eqref{DeltaL-CP-even-L} account for all the possible pure Higgs interactions, with the $p^2$ and $p^4$--operators $\cP_H$ and $\cP_{\Box H}$ respectively as
\be
\cP_H(h) = \frac{1}{2}\,(\derp_\mu h)^2\,\cF_H(h)\,,
\qquad
\cP_{\Box H}=\frac{1}{v^2}\,(\Box h)^2\,\cF_{\Box H}(h)\,.
\label{Opxih}
\ee

\nt More pure Higgs operators are possible, like $\cP_{DH}(h)$~\cite{Goldberger:2007zk, Contino:2011np} and $\cP_{\Delta H}$
\beq
\hspace{-0.22cm}\cP_{\small{DH}}(h)=\frac{\left((\derp_\mu h)^2\right)^2}{v^4}\cF_{\small{DH}}(h), \quad 
\cP_{\small{\Delta H}}= \frac{(\derp_\mu h)^2\Box h}{v^3}\cF_{\small{\Delta H}}(h)
\eeq
\nt not listed here as they are beyond the scope of this work.
The $\cG$--invariant CP--conserving effective chiral operators listed in Eq.~\eqref{CP-even-basis} and accounting for pure gauge and gauge-$h$ interactions, belong to three major categories: 

\begin{itemize}

\item[a)] $\{\cP_{1-3,\chi}(h),\,\cP_{6,\chi}(h),\,\cP_{11-14,\chi}(h),\,\cP_{23-24,\chi}(h)\}$ and $\cP_{26,\chi}(h)$ resulting from a direct extension of the original Appelquist-Longhitano chiral Higgsless basis already considered in Refs.~\cite{Appelquist:1980vg,Longhitano:1980iz,Longhitano:1980tm,
Feruglio:1992wf,Appelquist:1993ka}, coupled to the light Higgs $\cF(h)$ insertions, and after applying the discrete parity $P_{LR}$.

\item[b)] $\{\cP_{9,\chi}(h),\,\cP_{15-16,\chi}(h)\}$ containing the contraction $\cD_\mu \VLchimuu$ and no derivative couplings from $\cF(h)$. 

\item[c)] $\{\cP_{4-5,\chi}(h),\,\cP_{7-8,\chi}(h),\,\cP_{10,\chi}(h),\,\cP_{17-22,\chi}(h)\}$ and $\cP_{25,\chi}(h)$ with one or two derivative couplings from $\cF(h)$.

\end{itemize}

\nt It can be realized that the number of independent operators in the non--linear expansion turns out to be larger than for the analogous basis in the linear expansion, a generic feature when comparing both type of effective Lagrangians~\cite{Brivio:2013pma,Brivio:2014pfa}. The basis is also larger than that for the chiral expansions developed in the past for the case of a very heavy Higgs particle~\cite{Appelquist:1980vg,Longhitano:1980iz,Longhitano:1980tm,Appelquist:1993ka}, as:

\begin{itemize}

\item[i)] Terms which in the absence of the $\cF_i(h)$ functions were shown to be equivalent via total derivatives, are now independent.

\item[ii)] New terms including derivatives of $h$ appear.

\end{itemize}

\nt It is worth to comment that the LO custodial conserving $p^2$--structure $\Tr\left(\VLLmuu\,\VLLmud\right)$ together with the custodial breaking $p^2$--term $\left(\Tr\left(\TLL\,\VLLmuu\right)\right)^2$ (as well as the SM Yukawa terms not treated in here), can be straightforwardly extended in a more general manner by coupling them to the corresponding light Higgs dependence functions at the LO $p^2$--Lagrangian in Eq.~\eqref{LLO}, as
\be
\begin{aligned}
&\LL_0\,\supset \,\frac{1}{2} (\partial_\mu h) (\partial^\mu h) \,\left(1+c_H\,\cF_H(h)\right) 
   \,-\, V(h) \, + \, 
\\
& \hspace{-0.25cm}- \frac{\fL^2}{4}\Tr\left(\left(\VLLmuu\right)^2\right)\cF_{C,L}(h) + c_{T,L}\frac{\fL^2}{4}\Big(\Tr\left(\TLL\VLLmuu\right)\Big)^2\cF_{T,L}(h)
\end{aligned}
\label{Lh}
\ee

\nt where  $\cF_H(h)$, $\cF_{C,L}(h)$ and $\cF_{T,L}(h)$ are defined similarly as in Eq.~\eqref{F}. Particularly for $\cF_{C,L}(h)$, specific forms, alike in Eq.~\eqref{F}, were already provided in the literature~\cite{Contino:2010mh,Azatov:2012bz}. Constants $a$ and $b$ in Eq.~\eqref{F}  are model-dependent parameters, and specifically, $a$ and $c_{T,L}$ turn out to be constrained from electroweak precision tests as $0.7\lesssim a\lesssim1.2$ \cite{Azatov:2012qz} and $-1.7\times10^{-3}<c_{T,L}<1.9\times10^{-3}$ 
\cite{Giudice:2007fh} at $95\%$ CL.

The generalized LO $p^2$--Lagrangian $\LL_0$ of Eq.~\eqref{Lh} is useful in fact for describing an extended class of ``Higgs'' models:

\begin{itemize}

\item Mimicking the SM  scenario with a linear Higgs  sector after neglecting higher $h$-powers, and if $\mean{h}=v$, $a=b=1$.

\item Technicolor-like ansatz (for $\fL\sim v$ and omitting all terms in $h$) and intermediate situations with a light scalar $h$ from composite/holographic Higgs models \cite{Dimopoulos:1981xc,Kaplan:1983fs,Kaplan:1983sm,Banks:1984gj,
Georgi:1984ef,Georgi:1984af,Dugan:1984hq,Agashe:2004rs,Contino:2006qr,
Gripaios:2009pe} (in general for $\fL\ne v$).

\item Dilaton-like scalar frameworks \cite{Goldberger:2007zk,Halyo:1991pc,Vecchi:2010gj,Campbell:2011iw,
Matsuzaki:2012mk,Chacko:2012vm,Bellazzini:2012vz} (for $\fL\sim v$), where the dilaton participates to the electroweak symmetry breaking.

\end{itemize}

\nt Despite the usefulness of the Lagrangian $\LL_0$, it will be assumed for the analysis below the LO $p^2$--Lagrangian of Eq.~\eqref{LLO} henceforth. 

The connection of the non--linear framework analysed in here to the effective linear scenarios explicitly implementing the SM Higgs doublet, has been done for the case of the chiral Lagrangian $\LL_\text{chiral} = \LL_0
\,+\,\Delta \LL_{\text{CP},L}$ through Refs.~\cite{Alonso:2012px,Brivio:2013pma}, where all the operators contained in  Eq.~\eqref{DeltaL-CP-even-L} were respectively weighted by powers of $\xi=v^2/\fL^2$, in order to keep track of their corresponding operator siblings in the linear side. In fact, operators in the LO $p^2$--Lagrangian of Eq.~\eqref{LLO}, and those in the first line of Eq.~\eqref{DeltaL-CP-even-L}, as well as $\cP_{1-5,L}(h)$, had been already 
pointed out in the analysis of the linear--non linear connection of the SILH framework~\cite{Giudice:2007fh,Buchalla:2014eca}. Indeed, for the $\xi$--small limit, all the operators weighted by $\xi^{n\geq2}$ are negligible and the resulting Lagrangian has a correspondence with the SILH treatment. For the assumed non--linear scenario in this work, such linking between both of the EFT sides leads to 
account for the corresponding left--right symmetric extension of the effective linear approaches and it is beyond the scope of this paper.

On the other hand, a remarkable feature arises concerning the $P_{LR}$--symmetry for all the Lagrangian given so far. As it was commented in Section~\ref{P-LR-symmetry}, in the context of a general effective $SO(5)/SO(4)$ composite Higgs model scenario~\cite{Contino:2011np}, $P_{LR}$ was shown to be an accidental symmetry up to $p^2$--order and broken by several $p^4$--operators. Exactly the same properties are shared by the non--linear EW bosonic $\cG$-invariant scenario studied in this work, as it is suspected from the fact that $SU(2)_L \otimes SU(2)_R \sim SO(4)$. Indeed, $p^2$--operators in the LO Lagrangian $\LL_0\,+\,\LL_{0,R}
\,+\,\LL_{0,LR}$ described through Eqs.~\eqref{LLO}-\eqref{LLO-Left-Right}, explicitly exhibit $P_{LR}$ as an accidental symmetry of the approach. At higher momentum order in the Lagrangian expansion, the $p^4$--operators encoded in $\Delta \LL_{\text{CP}}$ through the Eqs.~\eqref{DeltaL-CP-even}--\eqref{CP-even-basis} and Eq.~\eqref{Opxih}, do not break $P_{LR}$ either. As soon as the 
$p^4$--operators made of mixed left and right--handed covariant structures are called in, non-zero contributions triggering the breaking of $P_{LR}$ will appear in the scenario as it will be seen a posteriori.

Up to now all the possible CP--invariant non--linear operators allowed by the local symmetry $\cG$ have been encoded up to the $p^4$--order in the first four pieces of $\LL_\text{chiral}$, i.e. in 
$\LL_0\,+\,\LL_{0,R}\,+\,\LL_{0,LR}\,+\Delta \LL_{\text{CP}}$ through Eqs.~\eqref{LLO}-\eqref{CP-even-basis} and Eq.~\eqref{Opxih}. In the following section the $SU(2)_L-SU(2)_R$ interplay between both of the symmetries is faced by accounting for all the possible mixed left--right symmetric interactions and up to the $p^4$--order via the remaining piece in $\LL_\text{chiral}$ Eq.~\eqref{Lchiral}, i.e. $\Delta\LL_{\text{CP},LR}$.


\boldmath
\subsection{$SU(2)_L-SU(2)_R$ interplay: $\Delta\LL_{\text{CP},LR}$}
\unboldmath

\nt As it was explained in Section~\ref{Interplay}, the construction of operators mixing the left and right--handed covariant objects lead to employ the covariant objects $\VLLRmuut$, $\TLLRt$ and $\WWLRut$ given in Eqs.~\eqref{Vtilde}, \eqref{Ttilde} and \eqref{Wtilde} respectively. Armed with these building tools, the complete set of operators accounting for the mixing between $SU(2)_L$ and $SU(2)_R$ covariant structures, can be parametrized as
\be
\begin{aligned}
&\Delta\LL_{\text{CP},LR}=\\
&\sum_{i=\{W,C,T\}}c_{i,LR}\cP_{i,LR}(h)\,\, \,+\,\,\,\sum_{i=2,\,i\neq 4}^{26} c_{i(j),LR}\,\cP_{i(j),LR}(h)\,,
\end{aligned}
\label{DeltaL-CP-even-LR}
\ee
\nt where the index $j$ spans over all the possible operators that can be built up from each $\cP_{i,\chi}(h)$ in Eq.~\eqref{CP-even-basis}, and here labelled as $\cP_{i(j),LR}(h)$ (as well as their corresponding coefficients $c_{i(j),LR}$), whilst the first term in $\Delta\LL_{\text{CP},LR}$ encodes the operators
\be
\begin{aligned}
\brown{\cP_{W,\,LR}(h)}\,\,  &= \brown{-\frac{1}{2}\,\gL\,\gR\,\Tr\left(\WWLut\,\WWRdt\right)\,\cF_{W,\,LR}(h)}  \\  
\cP_{C,\,LR}(h)\,\,  &= - \frac{1}{2}\,\fL\,\fR\,\Tr\Big(\VLLmuut\VLRmudt\Big)\,\cF_{C,\,LR}(h) \\
\cP_{T,\,LR}(h)\,\,  &= \frac{1}{2}\,\fL\,\fR\,\Tr\Big(\TLLt\,\VLLmuut\Big)\,\Tr\Big(\TLRt\,\VLRmudt\Big)\cF_{T,\,LR}(h)  
\label{WT-LR}
\end{aligned}
\ee

\nt corresponding to the ``mixed" versions of $\{\cP_{W,\chi}(h),\,\cP_{C,\chi}(h),\,\cP_{T,\chi}(h)\}$ in Eq.~\eqref{GT}, with $\cP_{W,\,LR}(h)$ missing in~\cite{Zhang:2007xy,Wang:2008nk}. The complete set of operators $\cP_{i(j),LR}(h)$ in the second term of $\Delta\LL_{\text{CP},LR}$ are listed as:

\begin{widetext}
\beq
\hspace{-0.5cm}
\begin{aligned}
&\cP_{2(1)}(h) = i\,g' \,B_{\mu\nu} \Tr\left(\TLLt\left[\VLLmuut,\VLRnuut\right]\right)\,\cF_{2(1)}\,, \qquad\qquad
&&\brown{\cP_{16(4)}(h) = \Tr\left([\TLRt,\VLRnuut]\,\cD_\mu \VLLmuut\right)\,\Tr\left(\TLLt\,\VLLnudt\right)\,\cF_{16(4)}}\,,\\[2mm]
&\cP_{3(1)}(h)  = i\,\gL\,\Tr\left(\WWLut\left[\VLRmudt,\VLRnudt\right]\right)\,\cF_{3(1)}\,,\qquad\qquad
&&\brown{\cP_{16(5)}(h) = \Tr\left([\TLRt,\VLRnuut]\,\cD_\mu \VLLmuut\right)\,\Tr\left(\TLRt\,\VLRnudt\right)\,\cF_{16(5)}}\,,\\[2mm]
&\cP_{3(2)}(h)  = i\,\gR\,\Tr\left(\WWRut\left[\VLLmudt,\VLLnudt\right]\right)\,\cF_{3(2)}\,,\qquad\qquad
&&\brown{\cP_{16(6)}(h) = \Tr\left([\TLLt,\VLLnuut]\,\cD_\mu \VLRmuut\right)\,\Tr\left(\TLLt\,\VLLnudt\right)\,\cF_{16(6)}}\,,\\[2mm]
&\cP_{3(3)}(h)  = i\,\gL\,\Tr\left(\WWLut\left[\VLLmudt,\VLRnudt\right]\right)\,\cF_{3(3)}\,,\qquad\qquad
&&\cP_{17(1)}(h) = i\,\gL\,\Tr\left(\TLLt\,\WWLut\right)\,\Tr\left(\TLRt\,\VLRmudt\right)\,\derp_\nu \cF_{17(1)}\,, \\[2mm]
&\cP_{3(4)}(h)  = i\,\gR\,\Tr\left(\WWRut\left[\VLLmudt,\VLRnudt\right]\right)\,\cF_{3(4)}\,,\qquad\qquad
&&\cP_{17(2)}(h) = i\,\gR\,\Tr\left(\TLRt\,\WWRut\right)\,\Tr\left(\TLLt\,\VLLmudt\right)\,\derp_\nu \cF_{17(2)}\,, \\[2mm]
&\brown{\cP_{5(1)}(h) = i\,\gL\,\Tr\left(\WWLut\,\VLRmudt\right)\,\derp_\nu \cF_{5(1)}}\,,\qquad\qquad
&&\cP_{18(1)}(h) = \Tr\left(\TLLt\,[\VLLmuut,\VLLnuut]\right)\,\Tr\left(\TLRt\,\VLRmudt\right)\,\derp_\nu\cF_{18(1)}\,, \\[2mm]
&\brown{\cP_{5(2)}(h) = i\,\gR\,\Tr\left(\WWRut\,\VLLmudt\right)\,\derp_\nu \cF_{5(2)}}\,,\qquad\qquad
&&\cP_{18(2)}(h) = \Tr\left(\TLRt\,[\VLRmuut,\VLRnuut]\right)\Tr\left(\TLLt\,\VLLmudt\right)\derp_\nu\cF_{18(2)}\,,\\[2mm] 
\end{aligned}
\label{CP-even-left-right-basis-I}
\eeq
\end{widetext}

\begin{widetext}
\beq
\hspace{-0.5cm}
\begin{aligned}   
&\cP_{6(1)}(h) =\left(\Tr\left(\VLLmuut\,\VLRmudt\right)\right)^2\cF_{6(1)}\,,
&&\brown{\cP_{18(3)}(h) = \Tr\left(\TLLt\,[\VLLmuut,\VLRnuut]\right)\Tr\left(\TLLt\,\VLLmudt\right)\derp_\nu\cF_{18(3)}}\,,  
\\[2mm]  
&\cP_{6(2)}(h) =\Tr\left(\VLLmuut\,\VLLmudt\right)\,\Tr\left(\VLRnuut\,\VLRnudt\right)\cF_{6(2)}\,,
&&\brown{\cP_{18(4)}(h) = \Tr\left(\TLRt\,[\VLLmuut,\VLRnuut]\right)\Tr\left(\TLRt\,\VLRnudt\right)\derp_\mu\cF_{18(4)}}\,,\\[2mm] 
&\cP_{6(3)}(h) =\Tr\left(\VLLmuut\,\VLLmudt\right)\,\Tr\left(\VLLnuut\,\VLRnudt\right)\cF_{6(3)}\,,
&&\brown{\cP_{18(5)}(h) = \Tr\left(\TLLt\,[\VLLmuut,\VLRnuut]\right)\Tr\left(\TLRt\,\VLRmudt\right)\derp_\nu\cF_{18(5)}}\,,\\[2mm]  
&\cP_{6(4)}(h) =\Tr\left(\VLRmuut\,\VLRmudt\right)\,\Tr\left(\VLLnuut\,\VLRnudt\right)\cF_{6(4)}\,,
&&\brown{\cP_{18(6)}(h) = \Tr\left(\TLRt\,[\VLLmuut,\VLRnuut]\right)\Tr\left(\TLLt\,\VLLnudt\right)\derp_\mu\cF_{18(6)}}\,,\\[2mm]
&\brown{\cP_{7(1)}(h) =\Tr\left(\VLLmuut\,\VLRmudt\right)\,\derp_\nu\derp^\nu\cF_{7(1)}}\,,
&&\brown{\cP_{19(1)}(h) = \Tr\left(\TLLt\,\cD_\mu\VLLmuut\right)\Tr\left(\TLRt\,\VLRnuut\right)\derp_\nu \cF_{19(1)}}\,, 
\\[2mm]
&\brown{\cP_{8(1)}(h) =\Tr\left(\VLLmuut\,\VLRnuut\right)\,\derp_\mu\cF_{8(1)}\,\derp_\nu\cF'_{8(1)}}\,,
&&\brown{\cP_{19(2)}(h) = \Tr\left(\TLRt\,\cD_\mu\VLRmuut\right)\Tr\left(\TLLt\,\VLLnuut\right)\derp_\nu \cF_{19(2)}}\,, 
\\[2mm]
&\brown{\cP_{9(1)}(h) = \Tr\left(\cD_\mu\VLLmuut\,\cD_\nu\VLRnuut \right)\,\cF_{9(1)}} \,,
&&\brown{\cP_{20(1)}(h) = \Tr\left(\VLLmudt\,\VLRmuut\right)\,\derp_\nu\cF_{20(1)}\,\derp^\nu\cF'_{20(1)}}\,,\\[2mm]
&\brown{\cP_{10(1)}(h) =\Tr\left(\VLLnuut \,\cD_\mu\VLRmuut\right)\,\derp_\nu \cF_{10(1)}}\,,
&&\cP_{21(1)}(h) = \Tr\left(\TLLt\,\VLLmuut\right)\Tr\left(\TLRt\,\VLRmudt\right)\left(\derp_\nu\cF_{21(1)}\right)^2\,,\\[2mm]
&\brown{\cP_{10(2)}(h) =\Tr\left(\VLRnuut \,\cD_\mu\VLLmuut\right)\,\derp_\nu \cF_{10(2)}}\,,
&&\cP_{22(1)}(h) = \Tr\left(\TLLt\,\VLLmuut\right)\Tr\left(\TLRt\,\VLRnuut\right)\derp_\mu\cF_{22(1)}\derp_\nu\cF'_{22(1)}\,,\\[2mm]
&\cP_{11(1)}(h) = \left(\Tr\left(\VLLmuut\,\VLRnuut\right)\right)^2\,\cF_{11(1)}\,,
&&\cP_{23(1)}(h) = \Tr\left(\VLLmuut\,\VLLmudt\right)\,\left(\Tr\left(\TLRt\,\VLRnuut\right)\right)^2\, \cF_{23(1)}\,,\\[2mm]
&\cP_{11(2)}(h) = \Tr\left(\VLLmuut\,\VLLnuut\right)\Tr\left(\VLRmudt\,\VLRnudt\right)\,\cF_{11(2)}\,,
&&\cP_{23(2)}(h) = \Tr\left(\VLRmuut\,\VLRmudt\right)\,\left(\Tr\left(\TLLt\,\VLLnuut\right)\right)^2\, \cF_{23(2)}\,,\\[2mm]
&\cP_{11(3)}(h) = \Tr\left(\VLLmuut\,\VLLnuut\right)\Tr\left(\VLLmudt\,\VLRnudt\right)\,\cF_{11(3)}\,,
&&\cP_{23(3)}(h) = \Tr\left(\VLLmuut\,\VLLmudt\right)
\Tr\left(\TLLt\,\VLLnuut\right)\Tr\left(\TLRt\,\VLRnudt\right) \cF_{23(3)}\,,
\\[2mm]
&\cP_{11(4)}(h) = \Tr\left(\VLRmuut\,\VLRnuut\right)\Tr\left(\VLLmudt\,\VLRnudt\right)\,\cF_{11(4)}\,,
&&\cP_{23(4)}(h) = \Tr\left(\VLRmuut\,\VLRmudt\right)
\Tr\left(\TLLt\,\VLLnuut\right)\Tr\left(\TLRt\,\VLRnudt\right) \cF_{23(4)}\,,\\[2mm]
&\cP_{11(5)}(h) = \Tr\left(\VLLmuut\,\VLRnuut\right)\Tr\left(\VLRmudt\,\VLLnudt\right)\,\cF_{11(5)}\,,
&&\cP_{23(5)}(h) = \Tr\left(\VLLmuut\,\VLRmudt\right)\,\left(\Tr\left(\TLRt\,\VLRnuut\right)\right)^2\, \cF_{23(5)}\,,\\[2mm]
&\cP_{12(1)}(h) = \gL\,\gR\,\Tr\left(\TLLt\,\WWLut\right)\Tr\left(\TLRt\,\WWRdt\right)\,\cF_{12(1)}\,, 
&&\cP_{23(6)}(h) = \Tr\left(\VLLmuut\,\VLRmudt\right)\left(\Tr\left(\TLLt\,\VLLnuut\right)\right)^2 \cF_{23(6)}\,,\\[2mm] 
&\cP_{13(1)}(h) = i\,\gL\,\Tr\left(\TLLt\,\WWLut\right)\,\Tr\left(\TLRt\left[\VLRmudt,\VLRnudt\right]\right)\,\cF_{13(1)}\,,
&&\cP_{23(7)}(h) = \Tr\left(\VLLmuut\,\VLRmudt\right)
\Tr\left(\TLLt\,\VLLnuut\right)\Tr\left(\TLRt\,\VLRnudt\right) \cF_{23(7)}\,,\\[2mm] 
&\cP_{13(2)}(h) = i\,\gR\,\Tr\left(\TLRt\,\WWRut\right)\,\Tr\left(\TLLt\left[\VLLmudt,\VLLnudt\right]\right)\,\cF_{13(2)}\,,
&&\cP_{24(1)}(h) = \Tr\left(\VLLmuut\,\VLLnuut\right)\Tr\left(\TLRt\,\VLRmudt\right)\Tr\left(\TLRt\,\VLRnudt\right)\cF_{24(1)}\,,\\[2mm] 
&\cP_{13(3)}(h) = i\,\gL\,\Tr\left(\TLLt\,\WWLut\right)\,\Tr\left(\TLLt\left[\VLLmudt,\VLRnudt\right]\right)\,\cF_{13(3)}\,,
&&\cP_{24(2)}(h) = \Tr\left(\VLRmuut\,\VLRnuut\right)\Tr\left(\TLLt\,\VLLmudt\right)\Tr\left(\TLLt\,\VLLnudt\right)\cF_{24(2)}\,,\\[2mm] 
&\cP_{13(4)}(h) = i\,\gR\,\Tr\left(\TLRt\,\WWRut\right)\,\Tr\left(\TLRt\left[\VLLmudt,\VLRnudt\right]\right)\,\cF_{13(4)}\,,
&&\cP_{24(3)}(h) = \Tr\left(\VLLmuut\,\VLLnuut\right)\Tr\left(\TLLt\,\VLLmudt\right)\Tr\left(\TLRt\,\VLRnudt\right)\cF_{24(3)}\,,\\[2mm] 
&\cP_{14(1)}(h) = \gL\,\epsilon_{\mu\nu\rho\sigma}\,\Tr\left(\TLRt\VLRmuut\right)
\Tr\left(\VLLnuut\,\WWWLut\right)\,\cF_{14(1)}\,, 
&&\cP_{24(4)}(h) = \Tr\left(\VLRmuut\,\VLRnuut\right)\Tr\left(\TLLt\,\VLLmudt\right)\Tr\left(\TLRt\,\VLRnudt\right)\cF_{24(4)}\,,\\[2mm] 
&\cP_{14(2)}(h) = \gR\,\epsilon_{\mu\nu\rho\sigma}\,\Tr\left(\TLLt\VLLmuut\right)
\Tr\left(\VLRnuut\,\WWWRut\right)\,\cF_{14(2)}\,, 
&&\cP_{24(5)}(h) = \Tr\left(\VLLmuut\,\VLRnuut\right)\Tr\left(\TLRt\,\VLRmudt\right)\Tr\left(\TLRt\,\VLRnudt\right)\cF_{24(5)}\,,\\[2mm] 
&\cP_{14(3)}(h) = \gL\,\epsilon_{\mu\nu\rho\sigma}\,\Tr\left(\TLLt\VLLmuut\right)
\Tr\left(\VLRnuut\,\WWWLut\right)\,\cF_{14(3)}\,, 
&&\cP_{24(6)}(h) = \Tr\left(\VLLmuut\,\VLRnuut\right)\Tr\left(\TLLt\,\VLLmudt\right)\Tr\left(\TLLt\,\VLLnudt\right)\cF_{24(6)}\,,\\[2mm]  
&\cP_{14(4)}(h) = \gR\,\epsilon_{\mu\nu\rho\sigma}\,\Tr\left(\TLRt\VLRmuut\right)
\Tr\left(\VLLnuut\,\WWWRut\right)\,\cF_{14(4)}\,,
&&\cP_{24(7)}(h) = \Tr\left(\VLLmuut\,\VLRnuut\right)\Tr\left(\TLLt\,\VLLmudt\right)\Tr\left(\TLRt\,\VLRnudt\right)\cF_{24(7)}\,,\\[2mm] 
\end{aligned}
\label{CP-even-left-right-basis-II}
\eeq
\end{widetext}

\begin{widetext}
\beq
\hspace{-0.5cm}
\begin{aligned}   
&\cP_{14(5)}(h) = \gR\,\epsilon_{\mu\nu\rho\sigma}\,\Tr\left(\TLLt\VLLmuut\right)
\Tr\left(\VLLnuut\,\WWWRut\right)\,\cF_{14(5)}\,, 
&&\cP_{24(8)}(h) = \Tr\left(\VLLmuut\,\VLRnuut\right)\Tr\left(\TLRt\,\VLRmudt\right)\Tr\left(\TLLt\,\VLLnudt\right)\cF_{24(8)}\,,\\[2mm]  
&\cP_{14(6)}(h) = \gL\,\epsilon_{\mu\nu\rho\sigma}\,\Tr\left(\TLRt\VLRmuut\right)
\Tr\left(\VLRnuut\,\WWWLut\right)\,\cF_{14(6)}\,, 
&&\brown{\cP_{25(1)}(h)= \Tr\left(\TLLt\,\VLLmuut\right)\Tr\left(\TLRt\,\VLRmudt\right)\derp_\nu\derp^\nu\cF_{25(1)}}\,,\\[2mm]
&\brown{\cP_{15(1)}(h) = \Tr\left(\TLLt\,\cD_\mu\VLLmuut\right)\,\,\Tr\left(\TLRt\,\cD_\nu\VLRnuut\right)\,\cF_{15(1)}}\,, 
&&\cP_{26(1)}(h)=\left(\Tr\left(\TLLt\,\VLLmuut\right)\Tr\left(\TLRt\,\VLRmudt\right)\right)^2\cF_{26(1)}\,,\\[2mm] 
&\brown{\cP_{16(1)}(h) = \Tr\left([\TLLt,\VLLnuut]\,\cD_\mu \VLLmuut\right)\,\Tr\left(\TLRt\,\VLRnudt\right)\,\cF_{16(1)}}\,,
&&\cP_{26(2)}(h)=\left(\Tr\left(\TLLt\,\VLLmuut\right)\Tr\left(\TLRt\,\VLRnuut\right)\right)^2\cF_{26(2)}\,,\\[2mm] 
&\brown{\cP_{16(2)}(h) = \Tr\left([\TLRt,\VLRnuut]\,\cD_\mu \VLRmuut\right)\,\Tr\left(\TLLt\,\VLLnudt\right)\,\cF_{16(2)}}\,,
&&\cP_{26(3)}(h)=\Tr\left(\TLLt\,\VLLmuut\right)\Tr\left(\TLRt\,\VLRmudt\right)\left(\Tr\left(\TLLt\,\VLLnuut\right)\right)^2\cF_{26(3)}\,,\\[2mm] 
&\brown{\cP_{16(3)}(h) = \Tr\left([\TLLt,\VLLnuut]\,\cD_\mu \VLRmuut\right)\,\Tr\left(\TLRt\,\VLRnudt\right)\,\cF_{16(3)}}\,,
&&\cP_{26(4)}(h)=\Tr\left(\TLLt\,\VLLmuut\right)\Tr\left(\TLRt\,\VLRmudt\right)\left(\Tr\left(\TLRt\,\VLRnuut\right)\right)^2\cF_{26(4)}\,,\\[2mm]
\end{aligned}
\label{CP-even-left-right-basis-III}
\eeq
\end{widetext}

\nt where the explicit light Higgs dependence is implicitly assumed through all $\cF_i(h)$ (here omitted in all $\cF_i(h)$ for shortness), although it has explicitly been kept for the operators nomenclature in Eqs.~\eqref{DeltaL-CP-even-LR}-\eqref{CP-even-left-right-basis-III}. Suffix $LR$ in all $\cP_{i(j),LR}(h)$ and their corresponding $\cF_{i(j),LR}(h)$ has been omitted as well in Eqs.~\eqref{CP-even-left-right-basis-I}-\eqref{CP-even-left-right-basis-III}. Among the total 75 operators $\cP_{i(j),LR}(h)$ listed in Eqs.~\eqref{WT-LR}-\eqref{CP-even-left-right-basis-III}, 23 operators (highlighted in red color again) are missing in the left-right symmetric EW chiral treatment of Refs.~\cite{Zhang:2007xy,Wang:2008nk}.

\begin{table}[h] 
\centering
\renewcommand{\arraystretch}{2.2}
\begin{tabular}{||c|c||} 
\hline \hline
\multicolumn{2}{||c||}{\bf{$P_{LR}$ symmetry}} \\[-3mm]
\multicolumn{2}{||c||}{$\Longleftrightarrow$} \\
\hline\hline
\quad$\cP_{6(3)}(h)$ \quad\quad & \quad$\cP_{6(4)}(h)$ \quad\quad\\
\quad$\cP_{10(1)}(h)$ \quad\quad & \quad$\cP_{10(2)}(h)$ \quad\quad\\
\quad$\cP_{11(3)}(h)$ \quad\quad & \quad$\cP_{11(4)}(h)$ \quad\quad\\
\quad$\cP_{16(1,3,5)}(h)$ \quad\quad & \quad$\cP_{16(2,4,6)}(h)$ 
\quad\quad\\
\quad$\cP_{18(1)}(h)$ \quad\quad & \quad$\cP_{18(2)}(h)$ \quad\quad\\
\quad$\cP_{18(3,5)}(h)$ \quad\quad & \quad$-\cP_{18(4,6)}(h)$ \quad\quad\\
\quad$\cP_{19(1)}(h)$ \quad\quad & \quad$\cP_{19(2)}(h)$ \quad\quad\\
\quad$\cP_{23(1,3,5)}(h)$ \quad\quad & \quad$\cP_{23(2,4,6)}(h)$ 
\quad\quad\\
\quad$\cP_{24(1,3,5)}(h)$ \quad\quad & \quad$\cP_{24(2,4,6)}(h)$ 
\quad\quad\\[1mm]
\hline\hline
\end{tabular}
\caption{\sf $P_{LR}$--symmetry acting over a subset of operators from $\Delta\LL_{\text{CP},LR}$ in Eq.~\eqref{DeltaL-CP-even-LR}-\eqref{CP-even-left-right-basis-III}. The rest of the operators from $\Delta\LL_{\text{CP},LR}$ (in the limit $\gL$, $\gR$, $g'\rightarrow 0$) not listed here are $P_{LR}$--even. Operators $\cP_{18(3-6)}(h)$ are explicitly triggering the breaking of $P_{LR}$.}
\label{PLR-tranformation-properties-even}
\end{table}

\nt Through the operators tower in Eqs.~\eqref{CP-even-left-right-basis-I}-\eqref{CP-even-left-right-basis-III} the definition for $\cD_\mu \VLLRmuut$ follows a similar one as in Eq.~\eqref{U-transforming-objects}
\be
\cD_\mu \VLLRmuut\equiv \UHLR^\dagger\,\cD_\mu \VLLRmuu\,\UHLR\,,
\label{DVtilde-covariant-derivative}
\ee

\nt where $\DLL^\mu \VLLRnuu$ has been defined in Eq.~\eqref{DV-covariant-derivative}. Notice that operators $\cP_{9(1)}(h)$, $\cP_{10(1-2)}(h)$,  $\cP_{15(1)}(h)$,  $\cP_{16(4-6)}(h)$, $\cP_{19(1-2)}(h)$, and $\cP_{16(1-3)}(h)$ containing the contraction $\cD_\mu \VLLRmuu$, whereas $\cP_{5(1-2)}(h)$, $\cP_{18(3-6)}(h)$ and $\cP_{19(1-2)}(h)$ involving one derivative of $\cF(h)$, and $\cP_{7(1)}(h)$, $\cP_{20(1)}(h)$, and $\cP_{25(1)}(h)$ with double derivatives of $\cF(h)$, are not present in  Refs.~\cite{Zhang:2007xy,Wang:2008nk}, and are the resulting additional ones from the allowed $SU(2)_L$--$SU(2)_R$ interplay of $\Delta\LL_{\text{CP},LR}$, together with the $P_{LR}$--symmetry.

The interplaying $\cG$--invariant CP--conserving non--linear  operators and up to the $p^4$--order in the Lagrangian expansion, listed in Eq.~\eqref{CP-even-left-right-basis-I}-\eqref{CP-even-left-right-basis-III}, can be catalogued as: 

\begin{itemize}

\item[a)] $\cP_{2(1)}(h)$, $\cP_{3(1-4)}(h)$, $\cP_{6(1-4)}(h)$, $\cP_{11(1-5)}(h)$, $\cP_{12(1)}(h)$, $\cP_{13(1-4)}(h)$, $\cP_{14(1-6)}(h)$, $\cP_{23(1-7)}(h)$, $\cP_{24(1-8)}(h)$ and $\cP_{26(1-4)}(h)$ coming from a direct extension of the original Appelquist-Longhitano chiral Higgsless basis already considered in Refs.~\cite{Appelquist:1980vg,Longhitano:1980iz,Longhitano:1980tm,
Feruglio:1992wf,Appelquist:1993ka}, coupled to the light Higgs $\cF(h)$ insertions, and accounting for each one of the operators than can be built up mixing left and right--handed covariant objects, supplemented by applying the discrete parity $P_{LR}$ to bring more mixing operators.

\item[b)] $\cP_{9(1)}(h)$, $\cP_{15(1)}(h)$ and $\cP_{16(1-6)}(h)$ containing the contraction $\cD_\mu \VLchimuu$ and no derivative couplings from $\cF(h)$. 

\item[c)] $\cP_{5(1-2)}(h)$, $\cP_{7(1)}(h)$, $\cP_{8(1)}(h)$, $\cP_{10(1-2)}(h)$, $\cP_{17(1-2)}(h)$, $\cP_{18(1-6)}(h)$, $\cP_{19(1-2)}(h)$, $\cP_{20(1)}(h)$, $\cP_{21(1)}(h)$, $\cP_{22(1)}(h)$ and $\cP_{25(1)}(h)$ with one or two derivative couplings from $\cF(h)$.

\end{itemize}

\nt As it is was mentioned in Section~\ref{Interplay}, after doing the proper replacement in Eqs.~\eqref{WT-LR}-\eqref{CP-even-left-right-basis-III} as $\VLLRmuut \rightarrow X^\mu_{L(R)}$ and $\WWLRut \rightarrow \frac{1}{\gLR}\overline{W}^{\mu\nu}_{L(R)}$ (following notation in~\cite{Zhang:2007xy,Wang:2008nk}), and further replacing the matrices $\TLLRt$ according to Eq.~\eqref{Ttilde}, all the operators mixing the left and right-handed covariant objects in~\cite{Zhang:2007xy,Wang:2008nk} are recovered, but the red coloured ones in Eqs.~\eqref{WT-LR}-\eqref{CP-even-left-right-basis-III}, resulting as commented above from the inclusion of the covariant structures $\cD_\mu \VLchimuu$ and derivative couplings from $\cF(h)$, plus the action of $P_{LR}$.

Finally, the transformation properties under the parity symmetry $P_{LR}$ of some of the operators from $\Delta\LL_{\text{CP},LR}$ in Eq.~\eqref{DeltaL-CP-even-LR}-\eqref{CP-even-left-right-basis-III} are exhibited in the Table~\ref{PLR-tranformation-properties-even}, with the operators not gathered in there behaving as $P_{LR}$--even (for the vanishing limit $\gL$, $\gR$, $g'\rightarrow 0$). Compact notation $\cP_{i(j,k,l),LR}(h)$ in the left column stands for operators $\cP_{i(j),LR}(h)$, $\cP_{i(k),LR}(h)$, $\cP_{i(l),LR}(h)$ reflected to $\cP_{i(m),LR}(h)$, $\cP_{i(n),LR}(h)$ $\cP_{i(p),LR}(h)$ (or vice versa) respectively, and collected by the operator notation $\cP_{i(m,n,p),LR}(h)$ in the right column. As anticipated before, a set of $p^4$--operators, in this case $\cP_{18(3-6),LR}(h)$, triggers the breaking of $P_{LR}$ explicitly, as it was expected from the general composite Higgs models grounds~\cite{Contino:2011np,Alonso:2014wta}. It is worth to comment on the transformation properties under the action of $P_{LR}$ at the Lagrangian level. In fact, requiring $P_{LR}$--invariance for the $p^2$--Lagrangian $\LL_0\,+\,\LL_{0,R}
\,+\,\LL_{0,LR}$ (Eqs.~\eqref{LLO}-\eqref{LLO-Left-Right}) and in the limit $\gL$, $\gR$, $g'\rightarrow 0$, will lead to the degenerated case $\fL \equiv \fR$. Furthermore, and before gauging the symmetry, $P_{LR}$--invariance for the Lagrangian $\Delta\LL_{\text{CP}}$ will imply $c_{i,L}\equiv c_{i,R}$ for the operator coefficients in Eqs.~\eqref{DeltaL-CP-even-L}-\eqref{DeltaL-CP-even-R}, as well as $a_{i,L}\equiv a_{i,R}$, $b_{i,L}\equiv b_{i,R}$ for the coefficient in the polynomial definition of Eq.~\eqref{F} (and for the rest of the coefficients in its expansion). Likewise for the Lagrangian $\Delta\LL_{\text{CP},LR}$, where the coefficients $c_{i(j,k,l),LR}$ for each one of the operators in the left column of Table~\ref{PLR-tranformation-properties-even} will have to be equal to the corresponding coefficients $c_{i(m,n,p),LR}$ in the right column (for the specific case of operators $\cP_{18(j),LR}(h)$ it leads to $c_{18(3,5),LR}=-c_{18(4,6),LR}$), and similarly for the implied coefficients in the light Higgs function definition. In the most general cases, such restrictive situation is not fulfilled, even less after the $\cG$--gauging is turned on. Henceforth, non-degenerate scales and operators coefficients will be assumed in general, i.e. no $P_{LR}$--invariance at the Lagrangian level.

Some of the CP--conserving bosonic operators provided above can be directly translated into pure bosonic operators  plus fermionic-bosonic ones. Such connection can be established through the covariant derivative $\cD_\mu\,\VLchimuu$ and the corresponding equation of motion for the light Higgs as it is described in the following.


\vspace{2mm}
\subsection{Linking to fermionic operators}

\nt Some set of operators from $\Delta\LL_{\text{CP}}$ (Eq.~\eqref{CP-even-basis}) and $\Delta\LL_{\text{CP},LR}$ (Eqs.~\eqref{CP-even-left-right-basis-I}-\eqref{CP-even-left-right-basis-III}) are independent only for the non-vanishing fermion masses case. Indeed, those operators containing the covariant contraction $\DLL_\mu\,\VLchimuu$ can be linked to Yukawa couplings~\cite{Alonso:2012px,Brivio:2013pma}, through the implementation of the gauge field EOM and the Dirac equations. When fermion masses are neglected, those operators can be written in terms of the other operators in the basis $\Delta\LL_{\text{CP}}$ and $\Delta\LL_{\text{CP},LR}$. In addition, using the light
$h$--EOM, operators with two derivative couplings of $\cF(h)$ 
can be reduced to a combination of  bosonic operators plus fermionic-bosonic ones. In general, all those operators must be included to have a complete and independent bosonic basis.

Considering the LO Lagrangian $\LL_0\,+\,\LL_{0,R}\,+\,\LL_{0,LR}$ described along Eqs.~\eqref{LLO}-\eqref{LLO-Left-Right}
 the equations of motion for the strength gauge fields $W^{\mu,a}_{L(R)}$ and $B^\mu$, and for the light Higgs $h$, are correspondingly given by
\begin{widetext}
\beq
\begin{aligned}
&\Big(D_\mu\,\WWLRu\Big)^a\,\,+\,\,\frac{\delta\LL_{f-kinetic}}{\delta W^{a}_{\nu,L(R)}}\,\,+\,\,(1+c_{W,LR})\,\partial_\mu \Big[\Tr\left(\UHLR^\dagger\,\tau^a\,\UHLR\,\WWRLut\right)\Big] =\\ \\
&\phantom{\Big(D_\mu\,\WWLRu\Big)^a\,\,+\,\,} \frac{i}{4}\,\gLR\,\Big\{\fLR^2\,\Tr\left(\VLLRnuu\,\tau^a\right) +(1+c_{C,\,LR})\,\fL\fR\Tr\left(\UHLR^\dagger\,\tau^a\,\UHLR\,\VLRLnuut\right)\Big\}
\left(1+\frac{h}{\fL}\right)^2\,,\\
\label{W-EOM}
\end{aligned}
\eeq

\beq
\begin{aligned}
&\derp_\mu B^{\mu\nu}\,\,+\,\,\frac{\delta\LL_{f-kinetic}}{\delta B_{\nu}}=\\ \\
&\phantom{\derp_\mu B^{\mu\nu}\,\,}-\frac{i}{4}\,g'\,\Big\{\Big[\fL^2\,\Tr\Big(\TLL\,\VLLnuu\Big)+\fR^2\,\Tr\Big(\TLR\,\VLRnuu\Big)\Big] 
 +\,\fL\,\fR\,(1+c_{C,\,LR})\,\Big[\Tr\Big(\TLRt\,\VLLnuut\Big)+\Tr\Big(\TLLt\,\VLRnuut\Big)\Big]\Big\}\left(1+\frac{h}{\fL}\right)^2\,,
\label{B-EOM}
\end{aligned}
\eeq

\beq
\begin{aligned}
&\square h\,\,+\,\,\frac{\delta\LL_{Yukawa}}{\delta h}\,\,+\,\,\dfrac{\delta V(h)}{\delta h}=\\ \\
&\phantom{\square h\,\,}-\frac{1}{2\,\fL}\left[\fL^2\,\Tr\Big(\VLLmuu\,\VLLmud\Big)+\fR^2\,\Tr\Big(\VLRmuu\,\VLRmud\Big)\right]\left(1+\frac{h}{\fL}\right)\,\,+\,\,(1+c_{C,\,LR})\,\frac{\fR}{2}\,\Tr\Big(\VLLmuut\,\VLRmudt\Big)\left(1+\frac{h}{\fL}\right)\,,
\label{h-EOM}
\end{aligned}
\eeq

\end{widetext}

\nt where the fermion dependent part of the EOMs has been generically encoded in the second terms at the first lines of Eqs.~\eqref{W-EOM}-\eqref{B-EOM}-\eqref{h-EOM} respectively, as no explicit kinetic fermion terms nor Yukawa interactions were accounted by $\LL_\text{chiral}$ in~\eqref{Lchiral}. From Eqs.~\eqref{W-EOM}-\eqref{B-EOM} it is derived

\begin{widetext}
\beq
\begin{aligned}
&\left[\fR\left(\fR + (1+c_{C,\,LR})\,\fL\right)\Tr\Big(\TLR\,\cD_\mu\VLRmuu\Big)+\left(R \leftrightarrow L\right)\right]\left(1+\frac{h}{\fL}\right)^2=\\ \\
&-2\,\left[\fL\left(\fL + (1+c_{C,\,LR})\,\fR\right)\Tr\Big(\TLL\,\VLLmuu\Big)+\left(L \leftrightarrow R\right)\right]\,\frac{\partial_\mu h}{\fL}\left(1+\frac{h}{\fL}\right)\,\,+\,\,\frac{4\,i}{g'}\,\partial_\mu\Big(\frac{\delta\LL_{f-kinetic}}{\delta B_{\mu}}\Big)\,,\\ \\
\end{aligned}
\label{T-DV-R}
\eeq

\beq
\begin{aligned}
&\Big\{\fLR\,\Tr\Big(\tau^a\,\cD_\mu\VLLRmuu\Big) +(1+c_{C,\,LR})\,\fRL\,\Tr\left(\UHLR^\dagger\,\tau^a\,\UHLR\,\cD_\mu \VLRLmuut\right)\Big\}
\left(1+\frac{h}{\fL}\right)^2=\\ \\
&\Big\{-2\,\fLR\Tr\Big(\tau^a\,\VLLRmuu\Big)\,\frac{\partial_\mu h}{\fL} -(1+c_{C,\,LR})\, \fRL\,\Tr\left(\UHLR^\dagger\,\tau^a\,\UHLR\,\left[\VLLRmuut,\VLRLmudt\right]\right)\Big\}
\left(1+\frac{h}{\fL}\right) +\\ \\
&- 2\,(1+c_{C,\,LR})\,\fRL\,\Tr\left(\UHLR^\dagger\,\tau^a\,\UHLR\,\VLRLmuut\right)\,
\frac{\partial_\mu h}{\fL}\,\left(1+\frac{h}{\fL}\right) -\frac{4\,i}{\gLR\,\fLR}\,\partial_\mu\Big(\frac{\delta\LL_{f-kinetic}}{\delta W^{a}_{\mu,L(R)}}\Big)\,,\\ 
\end{aligned}
\label{tau-DV-R}
\eeq

\end{widetext}

\nt where the last terms in Eqs.~\eqref{T-DV-R}-\eqref{tau-DV-R} can be translated into Yukawa terms via implementation of the corresponding Dirac equations. As it can be seen from the relations above, operators containing the contraction $\cD_\mu\,\VLRmuu$ can be translated into fermionic-bosonic operators plus pure bosonic ones, with some of them containing the contraction $\cD_\mu\,\VLLmuu$ (from the second terms in the left hand side of Eqs.~\eqref{T-DV-R}-\eqref{tau-DV-R}). Furthermore, by using the light Higgs-EOM in Eq.~\eqref{h-EOM}, those operators with two derivative couplings of $\cF(h)$ can be also rewritten in terms of pure bosonic ones plus fermionic-bosonic ones. More specifically,

\begin{itemize}

\item For the massless fermion case, operators $\{\cP_{9,\,R}(h),\,\cP_{15-16,\,R}(h)\,\cP_{19,\,R}(h)\}$ with the contraction $\cD_\mu\,\VLRmuu$ in Eq.~\eqref{CP-even-basis}, can be traded by pure bosonic operators contained in $\Delta\LL_{\text{CP}}$ (Eq.~\eqref{CP-even-basis}), some of them with the structure $\cD_\mu\,\VLLmuu$, and therefore they can be disregarded from the final operator basis in $\Delta\LL_{\text{CP}}$.  Similar feature applies for the operators $\cP_{9(1)}(h)$, $\cP_{10(1)}(h)$, $\cP_{15(1)}(h)$, $\cP_{16(2-3)}(h)$, $\cP_{16(6)}(h)$ and $\cP_{19(2)}(h)$ from $\Delta\LL_{\text{CP},LR}$ (Eqs.~\eqref{CP-even-left-right-basis-I}-\eqref{CP-even-left-right-basis-III}). In general, for the massive fermion case, all the previous operators have to be retained in the final basis.

\item For the vanishing fermion case, operators $\{\cP_{7,\,R}(h),\,\cP_{25,\,R}(h)\}$ with double derivatives of $\cF(h)$ in Eq.~\eqref{CP-even-basis} can be rewritten in terms of bosonic operators, with some of them contained in $\Delta\LL_{\text{CP}}$ and others in $\Delta\LL_{\text{CP},LR}$, and thus can be disregarded from the final operator basis. Likewise, operators $\{\cP_{7(1)}(h),\,\cP_{25(1)}(h)\}$ from $\Delta\LL_{\text{CP},LR}$ (Eqs.~\eqref{CP-even-left-right-basis-I}-\eqref{CP-even-left-right-basis-III}) are rewritable in
terms of operators in $\Delta\LL_{\text{CP},LR}$. When fermion masses are switched on, all the previous operators are physical and have to be included in the final basis.

\end{itemize}


\subsection{Integrating-out heavy right handed fields}
\label{Integrating-out}

\nt It is possible to integrate out the right handed gauge fields from the physical spectrum via the EOMs in Eqs.~\eqref{W-EOM}-\eqref{h-EOM} through the relations 
\be
\VLRmuu\,\equiv\,-\epsilon\,\,\VLLmuu\,,\qquad
\text{with}\qquad\epsilon\equiv \frac{\fL}{\fR}\,(1+c_{C,\,LR})
\label{Gauge-field-EOM}
\ee

\nt that can be translated into the unitary gauge as
\be
\begin{aligned}
W_{\mu,\,R}^\pm &\quad\Rightarrow\quad -\frac{\gL}{\gR}\,\epsilon\,\,W_{\mu,\,L}^\pm\,,\\
W_{\mu ,R}^3&\quad\Rightarrow\quad \frac{g'}{\gR}\left(1+\epsilon\right)B_{\mu } -\frac{\gL}{\gR}\,\epsilon\,\,W_{\mu ,L}^3
\label{Gauge-field-EOM-unitary-gauge}
\end{aligned}
\ee

\nt By plugging back the Eq.~\eqref{Gauge-field-EOM} through~\eqref{DeltaL-CP-even-R}-\eqref{CP-even-basis} (for $\chi=R$) and~\eqref{DeltaL-CP-even-LR}-\eqref{CP-even-left-right-basis-III}, all  the right handed and left-right operators will collapse onto the left  ones, affecting the corresponding global coefficients $c_{i,L}$ in a generic manner as
\be
c_{i,L}\,\,\Longrightarrow \,\,\tilde{c}_{i,L} \,=\, c_{i,L} \,+\, \sum^{4}_{k=1}\epsilon^k\,\cF^{(k)}\left(c_{i,R},\,c_{i(j)},\,c_{l(m)}\right) 
\label{Coefficients-redefined}
\ee

\nt where the functions $\cF^{(k)}\left(c_{i,R},c_{i(j)},\,c_{l(m)}\right)$ will encode linear combinations on the coefficients $c_{i,R}$, $c_{i(j)}$ and additional mixing left-right operators via $c_{l(m)}$. All the functions in~\eqref{Coefficients-redefined} can be easily read from~\cite{Shu:2015cxm}. The contributions induced onto the left handed operators are suppressed by powers of the ratio $\fL/\fR$, being determined by the number of fields $\VLRmuu$ through each one of the right and left--right operators. Consequently, in the limiting case $\fL\ll \fR$ at low energies, it is realized that the set of non-linear operators 
\be
\{\cP_B,\,\cP_{C,L},\,\cP_{T,L},\,\cP_{1,L},\,\cP_{2,L},\,\cP_{4,L}\}\,
\label{Sensitive-operators}
\ee

\nt is sensitive, up to the $\cO(\epsilon)$-contributions, either from the right handed operators
\be
\{\cP_{C,R},\,\cP_{T,R},\,\cP_{W,R},\,\cP_{1,R},\,\cP_{12,R}\}\,
\label{Right-operators}
\ee

\nt or the mixing left--right set
\be
\{\cP_{C,LR},\,\cP_{T,LR},\,\cP_{W,LR},\,\cP_{\text{3(2)}},\,\cP_{\text{12(1)}},\,\cP_{\text{13(2)}},\,\cP_{\text{17(2)}}\}\,.
\label{Left-Right-operators}
\ee  

\nt This is relevant for the EWPT parameters $S$ and $T$, as they are sensitive to the effects from $\cP_{1,L}$ and $\cP_{T,L}$ respectively~\cite{Shu:2015cxm}. The tree-level contributions to the oblique parameters $S$ and $T$~\cite{Peskin:1990zt}, turn out to be
\be
\begin{aligned}
\alpha_\text{em}\,\Delta S =  2\,s_{2W}\,\alpha_{WB}-8\,e^2\,\tilde{c}_{1,L}\,,\qquad
\alpha_\text{em}\,\Delta T =  2\,\tilde{c}_{T,L}\,,
\label{S-T-parameters}
\end{aligned}
\ee

\nt with $\alpha_\text{em}$ the fine structure constant and the notation $s_{2W}\equiv \sin(2\,\theta_W)$. The coefficient $\alpha_{WB}$ and the redefined ones $\tilde{c}_{1,L}$ and $\tilde{c}_{T,L}$ are defined as
\be
\begin{aligned}
\alpha_{WB}\,&\equiv \frac{g'}{2 \gR}\left(1-2\,\frac{\gL}{\gR}\,\epsilon\right)\left(1+\epsilon\right)\,,\\
\tilde{c}_{1,L}&=c_{1,L}\,-\,\frac{1}{4}\,c_{W,LR}\,+\,c_{\text{12(1)}}\,,
\\
\tilde{c}_{T,L}&=c_{T,L}\,+\,c_{T,R}\,-\,2 c_{T,LR}\,.
\end{aligned}
\ee

\vspace*{1mm}
\nt Furthermore, the triple gauge--boson couplings (TGC) are also sensitive to the induced effects by integrating out the right handed fields. These couplings can be generically described through the customary parametrization~\cite{Hagiwara:1986vm} 
\begin{widetext}
\be
\hspace*{-0.3cm}
\begin{aligned}
\frac{\LL_{\text{TGV}}}{g_{WWV}} =& \,i\Bigg\{ 
g_1^V \Big( W^+_{\mu\nu} W^{- \, \mu} V^{\nu} - 
W^+_{\mu} V_{\nu} W^{- \, \mu\nu} \Big) 
   \,+\, \kappa_V W_\mu^+ W_\nu^- V^{\mu\nu}\, + \, \\  \\
&\phantom{- \,i\,\,\,} -  ig_5^V \lambda^{\mu\nu\rho\sigma}
\left(W_\mu^+\partial_\rho W^-_\nu-W_\nu^-\partial_\rho W^+_\mu\right)
V_\sigma \,+\,
 g_{6}^V \left(\derp_\mu W^{+\mu} W^{-\nu}-\derp_\mu W^{-\mu} W^{+\nu}\right)
V_\nu  \Bigg\}\,,
\label{TGV-Lagrangian}
\end{aligned}
\ee
\end{widetext}

\nt where $V \equiv \{\gamma, Z\}$ and $g_{WW\gamma} \equiv e$, $g_{WWZ} \equiv e\,c_W/s_W$, with $W^\pm_{\mu\nu}$ and $V_{\mu\nu}$ standing for the kinetic part of the implied gauge field strengths. The compact notation $c_W\equiv \cos\theta_W$ and $s_W\equiv \sin\theta_W$ is implicit. Electromagnetic gauge invariance requires $g_{1}^{\gamma} =1$ and $g_5^\gamma=0$, in consequence the CP-even TGC encoded in~\eqref{TGV-Lagrangian} depends in all generality on six
dimensionless couplings $g_1^{Z}$, $g_5^Z$, $g_{6}^{\gamma,Z}$ and
$\kappa_{\gamma,Z}$. Their SM values are $g_1^{Z}=\kappa_{\gamma}=
\kappa_Z=1$ and $g_5^Z=g_{6}^{\gamma}=g_{6}^Z=0$. Additionally, the couplings $g_{6}^V$ have been introduced to account for the contributions associated to the operators containing the
contraction $\DLL_\mu\VLLmuu$, with its corresponding $\partial_\mu\VLLmuu$--part vanishing only for on-shell gauge bosons. 
When fermion masses are neglected, such contraction can be disregarded. The set of TGC parametrized through $\LL_{\text{TGV}}$ in~\eqref{TGV-Lagrangian} are written up to the $\cO(\epsilon)$-contributions as

\be
\begin{aligned}
\mathit{g}_{\text{\textit{$1$}}}^Z &= 1-\frac{2 s_W^4}{c_{2W} s_{2W}}\,\alpha_{WB} + \\
&\phantom{=}+\frac{1}{2 c_{2W}}\left[\tilde{c}_{T,L}-4 e^2 \left(c_{12,L}-\frac{s_W^2 \tilde{c}_{1,L}}{c_W^2}\right)\right]-\frac{4 e^2 c_{3,L}}{s_{2W}^2}\,,\\ \\
\kappa _{\text{\textit{$\gamma $}}}& = 1+ \frac{c_W}{s_W}\, \alpha_{WB} + \\
&\phantom{=} -\frac{e^2}{s_W^2}\left(2 \tilde{c}_{1,L}+2 \tilde{c}_{2,L}+c_{3,L}+4 c_{12,L}+2 c_{13,L}\right)\,, \\ \\
\kappa _{\text{\textit{$Z$}}}&= 1 -\frac{s_{2W}}{2 c_{2W}}\,\alpha_{WB} +\frac{\tilde{c}_{T,L}}{2\,c_{2W}}+ e^2 \left(\frac{2 \tilde{c}_{1,L}}{c_{2W}}+\frac{2 \tilde{c}_{2,L}}{c_W^2}\right)+ \\
&\phantom{=}- \frac{e^2}{s_W^2}\left[\left(\frac{1}{c_{2W}}+3\right) c_{12,L}+c_{3,L}+2 c_{13,L}\right]\,, \\ \\
\mathit{g}_{\text{\textit{$5$}}}^{\text{\textit{$Z$}}}&= -\frac{4 e^2}{s_{2W}^2}\,c_{14,L}\,,\qquad
\mathit{g}_{\text{\textit{$6$}}}^{\text{\textit{$\gamma $}}}=\frac{e^2}{s_W^2}\, c_{9,L}\,,\\
\mathit{g}_{\text{\textit{$6$}}}^Z & =e^2 \left(\frac{4 c_{16,L}}{s_{2 W}^2}-\frac{c_{9,L}}{c_{W}^2}\right)\,,\\
\end{aligned}
\ee

\nt with 
\be
\tilde{c}_{2,L}=c_{2,L}+\frac{1}{2} \left(2 c_{\text{13(2)}}+c_{\text{3(2)}}\right)\,.
\ee

\nt Likewise, some pair gauge bosons--Higgs couplings will be affected too. In fact, the vertexes $\{F_{\mu\nu} F^{\mu\nu} h,\,Z_{\mu\nu} Z^{\mu\nu} h,\,F_{\mu\nu} Z^{\mu\nu} h,\,Z_\mu\,Z^{\mu\nu}\,\partial_\nu h,\,Z_\mu\, F^{\mu\nu}\,\partial_\nu h\}$, and $\{W_\mu^\dag  W^\mu h,\,Z_\mu Z^\mu h\}$ will depend of linear combinations of the operators in~\eqref{Sensitive-operators}. See~\cite{Shu:2015cxm} for further details on the implied phenomenology and the allowed ranges for the involved operator coefficients.

As it was pointed out in~\cite{Brivio:2013pma} concerning the bounds on the $hVV$ couplings, the operators set $\{\cP_{7,L},\,\cP_{9,L},\,\cP_{10,L}\}$ was not included since their physical impact on the observables considered is negligible, while the set $\{\cP_{12,L},\,\cP_{17,L},\,\cP_{19,L},\,\cP_{25,L}\}$ doesn't entail a relevant contribution for the non-linear realization of the dynamics. In addition, the operators set $\{\cP_{9,L},\,\cP_{10,L},\,\cP_{15,L},\,\cP_{16,L},\,\cP_{19-21,L}\}$ becomes redundant for the massless fermion case via EOM~\cite{Brivio:2013pma}, while the set $\{\cP_{8,L},\,\cP_{18,L},\,\cP_{20-22,L}\}$ doesn't contribute directly to any of the couplings listed previously. These remarks lead us to have finally 19 left--handed operators that can be disregarded, and then, an effective set of 12 left ops. = 31 (set in~\eqref{GT}+\eqref{CP-even-basis})\,\,- 19. From all these considerations it is concluded that a right handed gauge sector far above the EW scale will imply a hierarchical case with NP effects parametrized via a much smaller operator basis as the $\fL/\fR$--suppression would entail, and leaving us therefore with 24 operators in total\,\,=\,\,12 left ops.\,\,+\,\,5 right ops. (in~\eqref{Right-operators})\,\,+\,\,7 left--right ops (in~\eqref{Left-Right-operators}).

\vspace*{3mm}

\section{Conclusions}
\label{Conclusions}

\nt In the tantalizing and challenging high energy regimes reachable at the LHC and future colliders, an effective Lagrangian approach is in order to parametrize all its possible physical effects detectable at low energies. In this paper, and concerning only the bosonic gauge sector, the NP field content is dictated by the existence of spin--1 resonances sourced by the extension of the SM local gauge symmetry $\cG_{SM}=SU(2)_L\otimes U(1)_Y$ up to the larger local group 
$\cG=SU(2)_L\otimes SU(2)_R\otimes U(1)_{B-L}$, here described via
a non--linear EW scenario with a light dynamical Higgs, and up to the $p^4$-contributions in the Lagrangian expansion.

This paper completes the CP--conserving pure gauge and gauge-$h$ operator basis given in  Refs.~\cite{Zhang:2007xy,Wang:2008nk} in the context of left--right symmetric EW chiral models, generalizing thus the work done in Refs.~\cite{Appelquist:1980vg,Longhitano:1980iz,Longhitano:1980tm,
Feruglio:1992wf,Appelquist:1993ka} for the heavy Higgs chiral scenario, and extending as well Refs.~\cite{Alonso:2012px,Brivio:2013pma} for the light Higgs dynamical framework, to the case of a larger local gauge symmetry $\cG$ in the context of non--linear EW interactions coupled to a light Higgs particle. The CP--violating counterpart of this analysis has been studied in a separate work~\cite{Yepes:2015qwa}.

The analysis provided in this work may also be considered as a generic UV completion of the low energy non--linear treatments of Refs.~\cite{Appelquist:1980vg,Longhitano:1980iz,Longhitano:1980tm,
Feruglio:1992wf,Appelquist:1993ka}  and  Refs.~\cite{Alonso:2012px,Brivio:2013pma}, as long as the extended gauge field sector arises out from an energy regime higher than the EW scale. The physical effects induced by integrating out the right handed fields from the physical spectrum are analysed. The relevant set of operators have been identified at low energies, 24 operators in total\,\,=\,\,12 left ops.\,\,+\,\,5 right ops. (in~\eqref{Right-operators})\,\,+\,\,7 left--right ops (in~\eqref{Left-Right-operators}). More low energy effects from a higher energy gauge sector~\cite{Shu:2015cxm,Shu:2016exh,Yepes:2015qwa} could unveil the underlying NP playing a role in our nature, and likely will point towards a better understanding on the origin of the electroweak symmetry breaking mechanism.

\section*{Acknowledgements}

\nt The author of this paper acknowledges valuable and enlightening comments and feedback from R.~Alonso, I.~Brivio, M.~B.~Gavela, J.~Gonzalez-Fraile, R. Kunming, D.~Marzocca, L. Merlo, J. Sanz-Cillero and J. Shu. The author also acknowledges KITPC financial support during the completion of this work.

%

\providecommand{\href}[2]{#2}\begingroup\raggedright\endgroup

\end{document}